\documentclass[a4paper,11pt]{article}
\usepackage{jcappub}
\usepackage{array}
\usepackage{multirow}
\usepackage{soul}
\usepackage[dvipsnames]{xcolor}

\graphicspath{{figures/}}

\title{\boldmath $B$-sure. Part II. Scattering transforms as robustness test for tensor-to-scalar ratio detection from CMB observations}

\author[*,1,2,3]{Claudio Ranucci,\note[*]{Corresponding author.}}
\author[4]{S\'ebastien Pierre,}
\author[5]{L\'eo Vacher,}
\author[1,2,3]{Alessandro Carones,}
\author[1,2,3]{Nicoletta Krachmalnicoff}
\author[4]{Erwan Allys,}
\author[6]{Jean-Marc Delouis,}
\author[7,8]{Paolo Campeti,}
\author[9,10,11]{Giuseppe Puglisi,}
\author[1,2,3]{and Carlo Baccigalupi}

\affiliation[1]{\it International School for Advanced Studies (SISSA),\\Via Bonomea 265, 34136 Trieste, Italy}
\affiliation[2]{\it Istituto Nazionale di Fisica Nucleare (INFN) Sezione di Trieste,\\Via Valerio 2, 34127 Trieste, Italy}
\affiliation[3]{\it Institute for Fundamental Physics of the Universe (IFPU),\\Via Beirut 2, 34151 Trieste, Italy}
\affiliation[4]{\it Laboratoire de Physique de l'\'Ecole Normale Sup\'erieure, ENS, Universit\'e PSL, CNRS, Sorbonne Universit\'e, Universit\'e Paris Cit\'e, 75005 Paris, France}
\affiliation[5]{\it Laboratoire de Physique des 2 infinis Irène Joliot-Curie (IJCLab), Universit\'e Paris-Saclay, CNRS/IN2P3, 91405 Orsay, France}
\affiliation[6]{\it Laboratoire d'Oc\'eanographie Physique et Spatiale (LOPS), Univ. Brest, CNRS, Ifremer, IRD, 29200 Brest, France}
\affiliation[7]{\it Dipartimento di Fisica e Scienze della Terra, Università degli Studi di Ferrara,\\Via Saragat 1, 44122 Ferrara, Italy}
\affiliation[8]{\it Istituto Nazionale di Fisica Nucleare (INFN) Sezione di Ferrara,\\Via Saragat 1, 44122 Ferrara, Italy}
\affiliation[9]{\it Dipartimento di Fisica e Astronomia, Università degli Studi di Catania,\\Via S. Sofia 64, 95123 Catania, Italy}
\affiliation[10]{\it Istituto Nazionale di Fisica Nucleare (INFN) Sezione di Catania,\\Via S. Sofia 64, 95123 Catania, Italy}
\affiliation[11]{\it Istituto Nazionale di Astrofisica (INAF) - Osservatorio Astrofisico di Catania,\\Via S. Sofia 78, 95123 Catania, Italy}

\emailAdd{claudio.ranucci.cr@gmail.com}

\abstract{Galactic foregrounds represent a major contamination to the measurement of primordial $B$-modes from observations of the Cosmic Microwave Background polarisation. Even after the application of component separation algorithms, foreground residuals may potentially still bias the estimate of the tensor-to-scalar ratio $r$, causing a false detection. In this work, we present the methodology of a robustness test for the validation of an eventual detection of primordial $B$-modes, as obtained by a future, \emph{LiteBIRD}-like satellite experiment. The goal of the test is to identify the foreground residuals contamination by looking for non-Gaussian properties in the CMB $B$-modes map, recovered through blind component separation algorithms. We adopt scattering transforms (ST) as our summary statistics sensitive to the non-Gaussian features of foreground residuals and to their correlation with foregrounds tracer maps. We characterise and validate the methodology on realistic sky simulations with different levels of foregrounds complexity. The proposed test is able to identify a bias on the tensor-to-scalar ratio of $\gtrsim 10^{-3}$ in $\sim 90\%$ of our simulations, with this bias value being of the same order of the accuracy targeted by \emph{LiteBIRD}. Additionally, for our particular experimental configuration, the test is passed when the bias is lower than the sensitivity on the $r$ parameter, and no warning is raised. These results provide an important step forward in the development of statistical tools for validating future measurement of cosmological parameters, against foregrounds contamination.}


\begin{document}

\maketitle

\flushbottom

\section{Introduction}
\label{sec:intro}
One of the major goals of current and future microwave sky experiments is the measurement of the primordial $B$-mode signal contained in the Cosmic Microwave Background (CMB) polarisation. This parity-odd pattern is generally attributed to a stochastic background of gravitational waves (GWs) in the primordial plasma, generated during a phase of exponential expansion known as cosmic inflation \cite{Brout78, Starobinsky80, Guth81, Kamionkowski97, Hu97, Seljak_Zaldarriaga97_gw}. The amplitude of the $B$-mode signal is usually parametrised through the tensor-to-scalar ratio $r$, a quantity directly proportional to the energy scale of inflation \cite{Planck20_inflation}. At the moment, there is no direct evidence for primordial $B$-modes yet, with the current constraints being $r_{0.05} < 0.032$ \cite{Tristram22} and $r_{0.01} < 0.028$ \cite{Galloni23} (with a free-to-vary tensor spectral tilt) at 95\% confidence when evaluated at a pivot scale of $0.05$ or $0.01 \, \mathrm{Mpc}^{-1}$, respectively.

Several factors complicate the measurement of the primordial $B$-mode signal. One of these is the gravitational deflection of the background CMB photons by the cosmic large-scale structure, which creates coherent sub-degree distortions in the CMB known as CMB lensing \cite{Zaldarriaga_Seljak98_lensing, Lewis06_lensing}. Through this mechanism, a fraction of the parity-even CMB $E$-modes is transformed into parity-odd $B$-modes at intermediate and small scales. Lensing $B$-modes have already been measured by \emph{SPTpol} \cite{SPT15}, \emph{ACTpol} \cite{ACT17_lensing}, \emph{PolarBear} \cite{Polarbear14} and \emph{BICEP2/Keck} \cite{BICEP18} experiments.

A second significant challenge is posed by diffuse Galactic foregrounds. Synchrotron radiation and thermal emission from dust grains generate $B$-modes which are dominant over the cosmological signal at all scales and even at high galactic latitudes \cite{Planck16_foregrounds, Planck16_PIP_dust, Skalidis18, Planck20_compsep, Planck20_dust}. At the minimum of their emission, around 80 GHz, the Galactic foregrounds $B$-mode signal represents an effective tensor-to-scalar ratio with amplitude larger than the sensitivity targeted by future microwave experiments, even in the cleanest regions of the sky \cite{Krach16_fgs}. Component separation methods are thus vital to address foregrounds contamination. Here we mention two categories: 1) parametric-fitting methods \cite{Eriksen08, Stompor08}, which recover the CMB signal by fitting a model of the various sky components; and 2) ``blind'' methods \cite{Delabrouille09_nilc, Carones23_mcnilc}, which instead do not make assumptions on the spectral energy distribution (SED) of foreground emission.

Component separation algorithms are able to mitigate most of the foregrounds impact, but residual contamination in the reconstructed CMB signal may still be comparable in amplitude to the primordial $B$-modes to be measured, thus biasing any estimate of the tensor-to-scalar ratio. In recent years, several works have tackled this problem (e.g., \cite{Alonso17_forecasts, LiteBIRD23PTEP, Carones23_nilc, Wolz24}), considering the sensitivity improvement forecasted by current and future experiments as the \emph{Simons Observatory} (\emph{SO}) \cite{SO19, SO25} and \emph{LiteBIRD}\footnote{\emph{LiteBIRD}: Lite (Light) satellite for the study of $B$-mode polarization and Inflation from cosmic background Radiation Detection.} \cite{LiteBIRD23PTEP}. These works have shown how foreground residuals could bias a $r \sim 10^{-3}$ measurement by several standard deviations $\sigma$. If left untreated, this can erroneously lead to a false detection, with the outcome of the \emph{BICEP2} 2014 analysis \cite{BICEP14, BICEP_Planck15} being a clear example. Thus, it is crucial to develop methodologies able to identify such residual contamination in the data, in order to validate a potential detection of the tensor-to-scalar ratio. In this context, statistical tools sensitive to peculiar properties of Galactic foregrounds must be considered when developing such robustness tests.

Scattering transform (ST) statistics are a class of summary statistics for the study of non-Gaussian processes \cite{Mallat11, Bruna12}. Inspired by neural networks but not requiring any training steps, they exploit iterative wavelet convolutions and non-linear operations to characterise interactions between oriented spatial scales. Recently introduced in astrophysics \cite{Allys19, Allys20}, ST statistics have demonstrated their ability to efficiently characterise a variety of non-Gaussian processes, as for instance the interstellar medium \cite{Blancard20, Lei23}, the large-scale structures (LSSs) of the Universe \cite{Cheng20, Valogiannis22a, Valogiannis22b}, and the epoch of reionisation \cite{Hothi24}. ST coefficients can also be used to build very efficient generative models in a maximum entropy framework, both for physical fields \cite{Bruna18, Cheng24, Mousset24, Hothi26} and CMB systematics datasets \cite{Campeti25}. In turn, this can be exploited to build efficient statistical component separation and inverse problem algorithms, which have in part been applied to real astrophysical data \cite{Blancard20, Delouis22, Tsouros26, Pierre26}. 

The aim of this paper is to use ST coefficients as summary statistics to build a robustness test for an eventual detection of the tensor-to-scalar ratio from CMB observations, adopting the configuration of a \emph{LiteBIRD}-like experiment, which represent the instrument with the most possibilities to constrain $r$ for years to come. The goal of the test is to identify the presence of Galactic foreground residuals left in the component-separated CMB maps, being a signature of an imperfect CMB signal reconstruction, possibly leading to a bias on $r$. We exploit the ST coefficients sensitivity to two different features: 1) the non-Gaussian properties of the foregrounds residuals in the CMB maps, and 2) the correlation of residual structures with foregrounds tracers. This extends what has been done in the first part of this series \cite{Ranucci26}, where Minkowski functionals (MFs) have been explored for the same purpose. We found that MFs are not sensitive enough for this purpose, failing to identify the bias $\delta(r) \sim 5 \times 10^{-3}$ induced by foreground residuals. ST statistics demonstrate instead to be very powerful in this context, being able to identify a $\delta(r) \sim 10^{-3}$ bias on the measurement of primordial $B$-modes in $\gtrsim 90\%$ of the simulations, following a general methodology applicable to different experiments and map-based component separation algorithms.

The paper is organised as follows. In section \ref{sec:method} we explain the methodology of the robustness test and the setup for its characterisation, and we also introduce the ST statistics (section \ref{sec:st}); results are reported and discussed in section \ref{sec:results}; finally, in section \ref{sec:conclusions} we summarise our findings and provide future prospects, with additional details provided in the appendices.

\section{Methodology and experimental setup}
\label{sec:method}
\begin{figure}
    \centering
    \includegraphics[width=1.\textwidth]{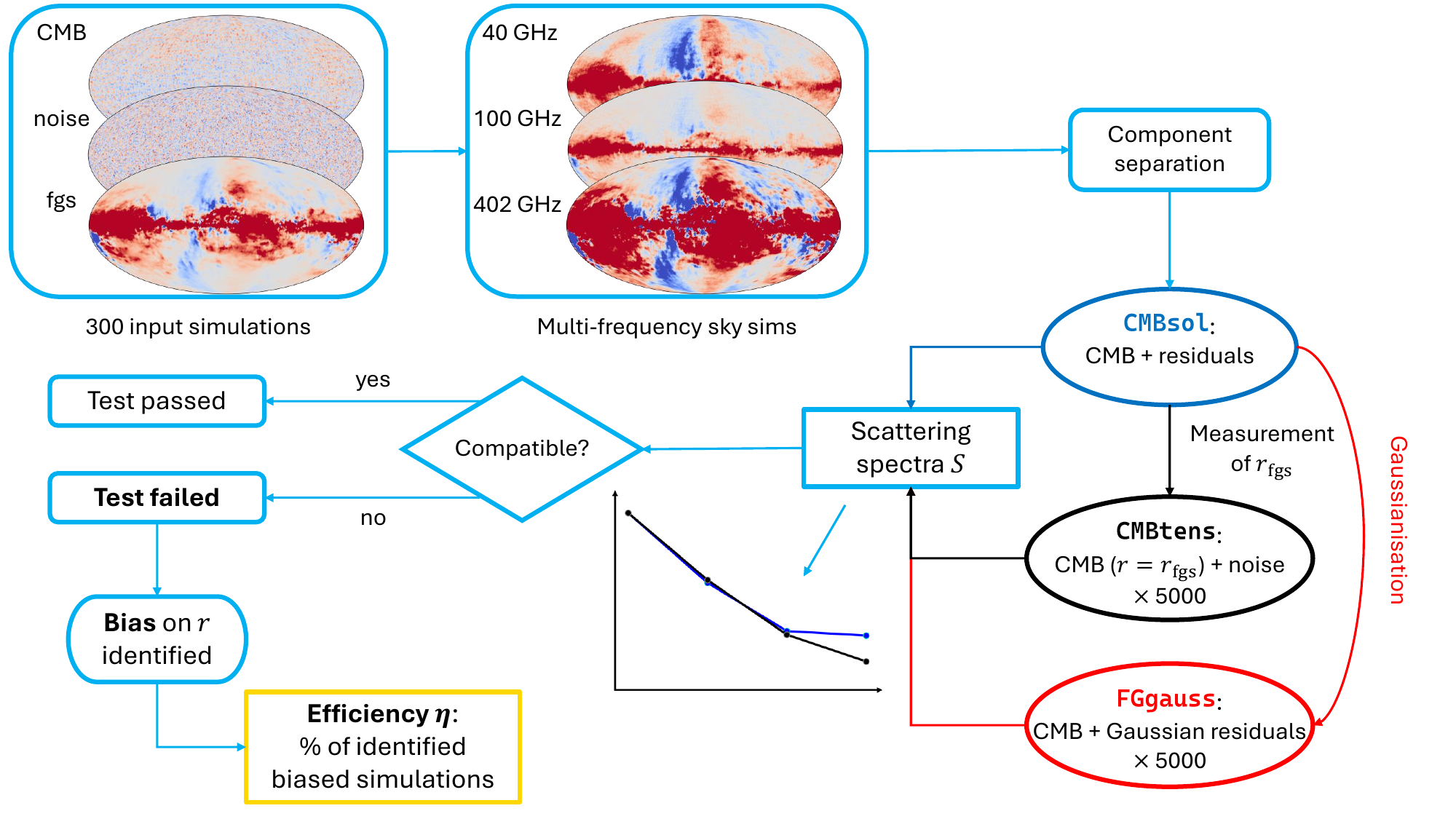}
    \caption{Flowchart illustrating the methodology of the ST-based robustness test, showing the various steps followed for its characterisation.}
    \label{fig:flowchart}
\end{figure}

In this section, we lay down the methodology adopted for the construction and characterisation of the robustness test. We intentionally omit some details and discussion points that have already been treated in the previous part of the series, to which we refer to \cite{Ranucci26} for a more complete description of the setup, products, and analysis choices adopted. Appendices of the present work also contain supplementary technical information on specific topics. To help a smoother reading, figure \ref{fig:flowchart} provides a schematic illustration of the pipeline underlying the robustness test.

\subsection{Input simulations}
\label{sec:sims}
Maps generation, manipulation, and analysis are carried out adopting the \texttt{HEALPix}\footnote{\href{https://healpix.jpl.nasa.gov/}{https://healpix.jpl.nasa.gov/}} \cite{healpix} pixelisation scheme by means of the \texttt{healpy}\footnote{\href{https://healpy.readthedocs.io/en/latest/}{https://healpy.readthedocs.io/en/latest/}} package \cite{healpy}. For all our simulations, the pixel resolution is defined by $N_\text{side} = 64$, corresponding to an angular resolution of $\sim 55$ arcmin, with the maximum multipole considered being $\ell_\text{max} = 3 N_\text{side} - 1 = 191$.

We simulate Stokes $Q$/$U$ sky maps as observed by a \emph{LiteBIRD}-like experiment, containing CMB, noise, and Galactic foregrounds components. \emph{LiteBIRD} specifications are taken from the pre-reformation forecast of the experiment (table 13 of \cite{LiteBIRD23PTEP}), planned to cover frequencies between 40 and 402 GHz. We generate two sets of 300 CMB Gaussian realisations from the angular power spectra of the best-fitting \emph{Planck} 2018 parameters \cite{Planck20_params}. In one set, $B$-modes are sourced by gravitational lensing only, with no primordial signal ($r_\text{in} = 0$); in the other, we also have an amplitude of primordial $B$-modes sourced by inflationary GWs ($r_\text{in} = 5 \times 10^{-3}$), compatible with current experimental upper limits. Noise maps are generated as Gaussian realisations of white noise at the different frequency channels. The $1/f$ component of the noise is neglected here, as it is expected to be fully suppressed by the presence of the half-wave plate \cite{LiteBIRD23PTEP}.

Galactic foregrounds are simulated with the Python Sky Model (\texttt{PySM})\footnote{\href{https://pysm3.readthedocs.io/en/latest/index.html}{https://pysm3.readthedocs.io/en/latest/index.html}} \cite{PySM, Zonca21_PySM, Panexp25_PySM}. We focus on thermal dust (\texttt{d}) and synchrotron (\texttt{s}) emissions, as they are the main contaminants of polarised observations. We adopt two different models of foreground emission: \texttt{d0s0} and \texttt{d10s5}. The former is a simplistic view of the sky that we use for the testing and benchmarking of the robustness test, while the latter is a more complex model closer to realistic observations. We refer to the \texttt{PySM} papers and documentation for more details. We generate full-sky simulations for each model, one per frequency channel.

In order to be ingested by the successive component separation algorithms, all the simulated maps are smoothed to a common angular resolution of $\text{FWHM}_\text{out} = 70.5 \, \mathrm{arcmin}$, corresponding to the largest beam considered (40 GHz channel). We notice that, technically, this angular resolution is not sufficiently large compared to the pixel size associated to the adopted $N_\text{side}$ (55 arcmin). The angular resolution should be at least 3 times the pixel resolution, while here is instead only 1.5 larger. However, we found this effect to have negligible impact, in any case impacting only the smaller scales. In this work we will mainly focus our attention on the larger scales, the most contaminated by foregrounds emission.

We end up with sets of sky simulations which include CMB, noise, and foregrounds, with 300 different realisations of CMB (with and without $r$) and noise for each considered \texttt{PySM} model. The $Q/U$ simulations are then converted to $B$-mode maps through a full-sky harmonic transformation, and this will be our input dataset for component separation. In this work, the non-Gaussian features induced by the gravitational lensing of large-scale structures are not included in the CMB simulations, generated as Gaussian realisations. We verified that the statistics we adopt here are not sensitive enough to identify the impact of the lensing non-Gaussian contribution, which is anyway expected to be important on the smaller scales.

\subsection{Tensor-to-scalar ratio estimation}
\label{sec:r_estimate}
In this work, we adopt blind algorithms as our component separation methods to recover CMB $B$-modes from multi-frequency simulations of the sky. In particular, we use two implementations: Needlet Internal Linear Combination (NILC) and Multi-Clustering NILC (MC-NILC). NILC has already been applied in the data analysis of past CMB experiments (e.g., \emph{WMAP} \cite{Basak13_nilc_pol}, \emph{Planck} \cite{Planck20_compsep}) and both NILC and MC-NILC will be used for current and future CMB surveys (\emph{SO} \cite{SO19, Wolz24}, \emph{LiteBIRD} \cite{LiteBIRD23PTEP}). We refer to \cite{Carones23_nilc, Carones23_mcnilc} for complete discussion of the two techniques, while some additional details are reported in appendix \ref{sec:compsep}. Both NILC and MC-NILC methods are applied as implemented in the \texttt{BROOM}\footnote{\href{https://github.com/alecarones/broom}{https://github.com/alecarones/broom}} \texttt{python} package \cite{Carones26_broom}.

\begin{figure}
    \centering
    \includegraphics[width=1.\textwidth]{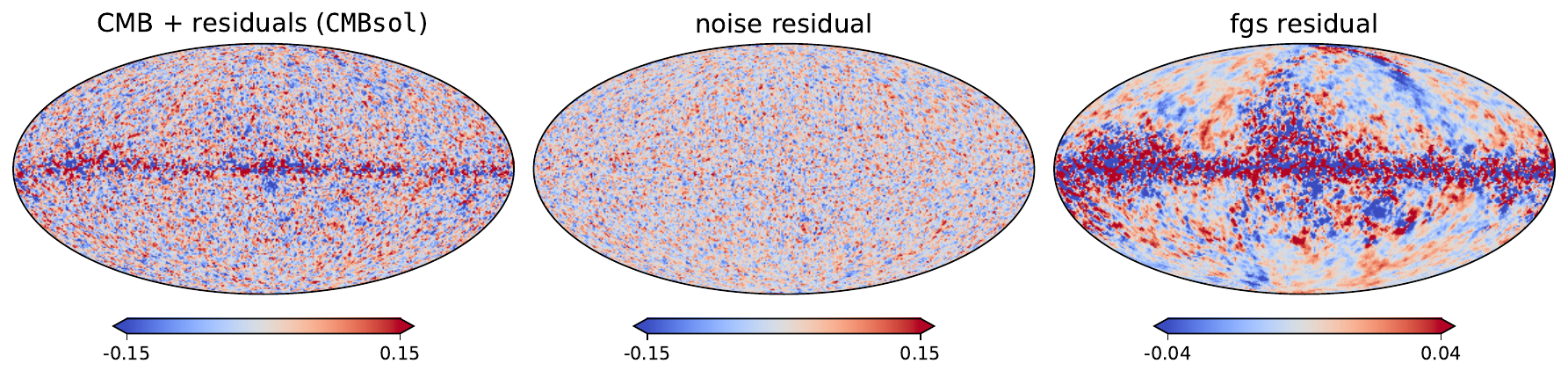}
    \caption{Component separation results, for one simulation. \emph{Left:} recovered CMB $B$-mode signal (\texttt{CMBsol}), containing residual contamination. \emph{Middle:} noise residuals. \emph{Right:} foreground residuals. Outputs are shown for the MC-NILC method, applied on the \texttt{d10s5} model. Units are $\mu \mathrm{K_{CMB}}$.}
    \label{fig:compsep_maps}
\end{figure}

In general, the outcome of a map-based component separation algorithm is a single CMB ``solution'', which we denote \texttt{CMBsol}, containing the cleaned CMB signal and some residual contamination from noise and foregrounds, whose level depend on the performances of the cleaning method. In a real experiment, the isolated contributions of these residuals are not accessible. Since in this analysis we are working with simulated datasets, we have complete control over the separated sky components, as we independently generated CMB, noise, and foregrounds. Component separation weights can be applied to the input noise and foregrounds maps to obtain maps of the residuals, useful for pipeline validation and for the characterisation of the robustness test. Figure \ref{fig:compsep_maps} shows an example of the results obtained from the application of the MC-NILC component separation on the \texttt{d10s5} model.

We notice that, in real data, potential correlations between noise and foregrounds maps can arise due to the map-making process, for instance because of bandpass differences between detectors. A simple sum of independent components is not able to replicate these small correlations, requiring a full simulation of the map-making procedure on the full signal. This may leave some residual foreground contribution, with a more Gaussian statistics due to the mixing with noise. We do not attempt to include this additional effect in our simulations, since it would only enhance the non-Gaussian features of the foreground residuals, improving the final performance of the robustness test.

To infer the tensor-to-scalar ratio from the \texttt{CMBsol} map, we adopt an inverse-Wishart distribution as our likelihood \cite{Hamimeche_Lewis09, Gerbino20_likelihood}:
\begin{equation}
    - \ln \mathcal{L} \left( C_\ell^\text{obs} | r \right) = \frac{1}{2} \sum_{\ell} (2\ell + 1) f_\text{sky} \left[ \frac{C_\ell^\text{obs}}{C_\ell^\text{th}(r)} + \ln C_\ell^\text{th}(r) - \frac{2\ell - 1}{2\ell + 1} \ln C_\ell^\text{obs} \right] \, .
    \label{eq:likelihood}
\end{equation}
where $C_\ell$ denotes the usual angular power spectrum, $C_\ell^\text{obs}$ is the ``observed'' power spectrum evaluated on \texttt{CMBsol}, containing CMB, noise and foreground residuals, and $C_\ell^\text{th}$ is the theoretical $BB$ spectrum built as
\begin{equation}
    C_\ell^\text{th}(r) = C_\ell^\text{lens} + \left\langle C_\ell^\text{nres} \right\rangle + r \cdot C_\ell^{r=1} \, .
    \label{eq:cl_th}
\end{equation}
In the above equation, $C_\ell^\text{lens}$ is the $B$-mode power spectrum induced by gravitational lensing, $C_\ell^{r=1}$ is the theoretical $B$-mode power spectrum sourced solely by tensor perturbations with $r = 1$, and $\langle C_\ell^\text{nres} \rangle$ is the average of the spectra computed on all the other 299 noise residuals maps obtained from the component separation. In a real experiment, the latter would be replaced by the average over a sample of simulations apt at the reproduction of the noise properties of the instrument.

Angular power spectra are computed with the \texttt{anafast} routine as implemented in the \texttt{healpy} package \cite{healpix, healpy}. To avoid strong contamination from the Galactic plane, a mask\footnote{Available at the \href{https://pla.esac.esa.int/\#maps}{Planck Legacy Archive}.} is applied to the maps before power spectra computation, retaining a sky fraction of $f_\text{sky} \simeq 60\%$ (shown as the gray region of the maps of figure \ref{fig:maps_test}). The \texttt{anafast} routine does not account for correlations among multipoles induced by the presence of a mask \cite{Lewis01_masked}. As discussed in \cite{LiteBIRD23PTEP}, this approximation has a negligible impact on the power spectra estimation of foreground and noise residuals over large sky fractions, as considered in this work.

\begin{figure}
    \centering
    \includegraphics[width=1\textwidth]{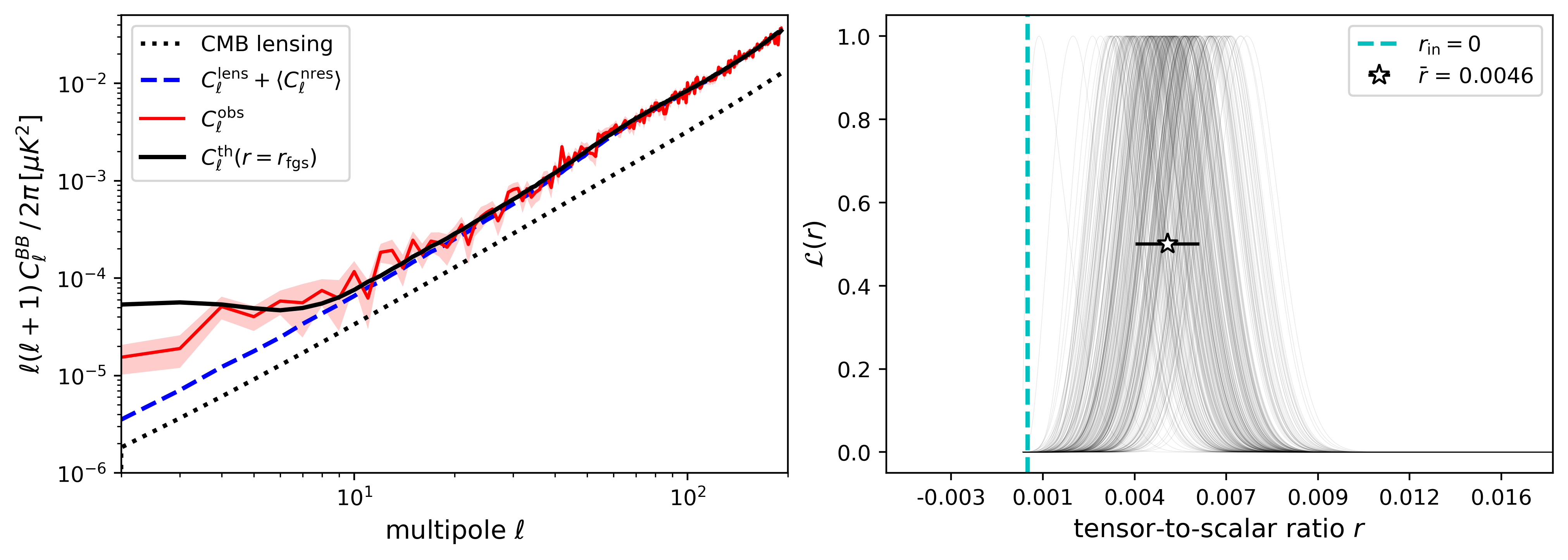}
    \caption{\emph{Left panel}: angular power spectra for one simulation, with $C_\ell^\text{th} (r = 0)$ in dashed blue, $C_\ell^\text{obs}$ in red solid, $C_\ell^\text{lens}$ in dotted black, and best fit in black solid. The shaded area represents the standard deviation of $C_\ell^\text{obs}$ across 300 realisations. \emph{Right panel}: normalised likelihood on the tensor-to-scalar ratio $r$ for the 300 simulations, with the input value indicated by the vertical dashed line, and the star reporting the mean value and the standard deviation over the simulations. In both panels, the considered scenario is MC-NILC - \texttt{d10s5} - $r_\text{in} = 0$.}
    \label{fig:spectra_likelihood}
\end{figure}

We take the peak value of the evaluated likelihood as the best fit tensor-to-scalar ratio $r_\text{fgs}$, where the $\textbf{fgs}$ label indicates that the (eventual) bias on the parameter is entirely due to the presence of foreground residuals, responsible for the excess power at large scales with respect to the CMB (lensing and primordial) signal. Figure \ref{fig:spectra_likelihood} shows the angular power spectra for one simulation, together with the likelihood evaluated for all the 300 simulations.

\subsection{Reference Gaussian simulations}
\label{sec:sims_gauss}
The goal of the proposed robustness test is to compare the statistical properties of the recovered CMB signal (\texttt{CMBsol}) with reference Gaussian simulations, going beyond the usual two-point statistics. Foreground residuals left by component separation are expected to show non-Gaussian properties and anisotropic structures which are not expected to be present in a field dominated by primordial CMB $B$-modes\footnote{We note the possibility for the CMB field to contain a small amount of primordial non-Gaussianity \cite{Bartolo04}. At the moment, we only have upper limits on them, being anyway far less impactful with respect to foregrounds non-Gaussianity. We neglect primordial non-Gaussianity in this work.}. Thus, we build two sets of Gaussian simulations dubbed \texttt{CMBtens} and \texttt{FGgauss}, which are co-added with the input CMB lensing maps and noise residuals obtained from component separation. These CMB and noise realisations are independent from the CMB and noise contained in \texttt{CMBsol}.

If the detected tensor-to-scalar ratio is of Galactic origin, then the \texttt{CMBsol} map and the Gaussian simulations should be statistically incompatible when observed through a higher-order statistics and fail the robustness test, since \texttt{CMBsol} contains anisotropic contamination. Vice versa, if no deviation is observed, it means that the methodology is not able to detect the foreground residuals non-Gaussian contribution. In the latter case, the test is passed and no flag is raised. In practice, we statistically compare the \texttt{CMBsol} map with two sets of Gaussian simulations:

\paragraph{\texorpdfstring{\texttt{CMBtens}}{CMBtens}} For each of the 300 \texttt{CMBsol} simulations, we estimated a tensor-to-scalar ratio $r_\text{fgs}$, eventually biased by foreground residuals. We use each inferred value to generate 5000 Gaussian realisations of $B$-mode maps sourced solely by primordial GWs ($C_\ell^{BB} = r_\text{fgs} \cdot C_\ell^{r = 1}$), smoothed to the resolution of \texttt{CMBsol}. We then co-add these simulations with independent realisations of CMB lensing and noise residuals maps. In general, the power spectrum of these \texttt{CMBtens} simulations is similar in shape and amplitude to the one of \texttt{CMBsol} maps; their CMB lensing and noise contributions are statistically the same, while the tensor contribution is fitted on the spectrum contaminated by foregrounds. However, it is possible that the two spectrum shapes do not match enough to pass a standard goodness-of-fit analysis, which can also be classified as a first robustness test on the $r$ estimate. A standard goodness-of-fit analysis has been detailed in the previous part of this series of papers \cite{Ranucci26}.

\paragraph{\texorpdfstring{\texttt{FGgauss}}{FGgauss}} In order to remove the mentioned differences at the power spectrum level and only characterise deviations from Gaussianity, we replace the non-Gaussian foreground residuals contained in \texttt{CMBsol} with a Gaussian version of them. Specifically, we compute the power spectrum (outside of a Galactic mask) of each of the 300 foreground residuals maps, and we use it to generate 5000 Gaussian realisations through the \texttt{synfast} routine of the \texttt{healpy} package. This means that we are treating the spectrum of a single realisation as the correct underlying theoretical spectrum, with the average spectrum of the Gaussian simulations converging to the non-Gaussian one. In this way, the \texttt{CMBsol} and \texttt{FGgauss} maps have the same power spectrum; the only difference between them is in the non-Gaussian properties. This is equivalent to having an optimal fit in a realistic experiment. We stress that, for real data, we would not have access to the component separation residuals and thus to the possibility of generating the \texttt{FGgauss} type of simulations\footnote{Even if we would still be able to generate Gaussian versions of the recovered \texttt{CMBsol} map.}. Nonetheless, they are useful here for a full characterisation of the robustness test in a controlled scenario. On the other hand, it is always possible to generate the \texttt{CMBtens} simulations, even in a real experiment.

\begin{figure}
    \centering
    \includegraphics[width=1.0\textwidth]{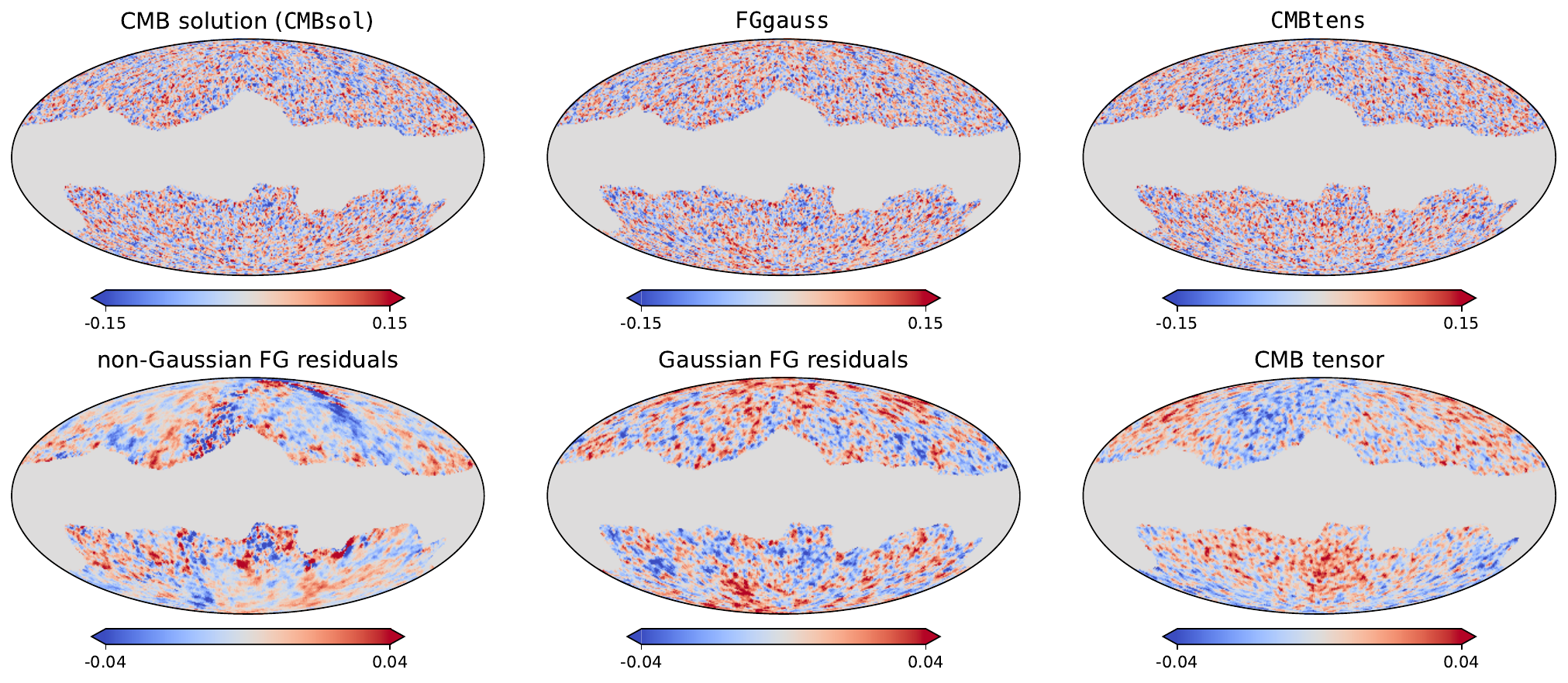}
    \caption{Example of maps used in the robustness test characterisation, for the MC-NILC - \texttt{d10s5} - $r_\text{in} = 0$ case. \emph{Upper row:} CMB solution (\texttt{CMBsol}) on the left, Gaussian simulations \texttt{FGgauss} and \texttt{CMBtens} in the middle and on the right, respectively. \emph{Lower row:} as the upper row, without CMB lensing and noise, with non-Gaussian (left) and Gaussian (middle) foreground residuals, and tensor perturbations (right).}
    \label{fig:maps_test}
\end{figure}

An example of the two sets of Gaussian reference simulations is shown in figure \ref{fig:maps_test}. We notice the strong similarity at the map level between the Gaussian and slightly non-Gaussian maps, since they are mainly dominated by the (Gaussian) noise component. Higher-order statistics are required to identify the morphological differences.

\subsection{Scattering transform statistics}
\label{sec:st}
As mentioned in section \ref{sec:intro}, we build our robustness test on Scattering transform (ST) statistics. Here we give a brief introduction to the concepts needed to discuss our results, referring to \cite{Mallat11, Bruna12, Allys19, Delouis22, Cheng24} for more complete presentations. ST are summary statistics for high-dimensional non-Gaussian signals, built from a cascade of convolutions with fixed filters (a wavelet transform), a modulus operation, and spatial averages or covariances. The filters, here Morlet wavelets, are defined analytically, rather than learned from data. A key advantage of ST over other higher-order statistics is their lower variance. Indeed, estimators based directly on moments of the field, such as the bi-spectrum, have a large variance dominated by outliers \cite{Welling05}, while using the modulus as the non-linearity instead yields statistics with a controlled variance. This is central to our purpose, where the non-Gaussian signal is faint and must be told apart from sampling variance.

The wavelet transforms are computed in pixel space on the sphere, by convolution with a family of dyadic Morlet wavelets $\psi_{(j, \theta)}$, obtained by dilating by a factor $2^j$ and rotating by a factor $\theta \cdot \pi/4$ a mother wavelet. In practice, we label $\lambda = (j, \theta)$ the oriented scales, and $\psi_\lambda$ the related wavelets. We probe $J = 6$ dyadic scales ($j_\text{min} = 0$ to $j_\text{max} = 5$) and $\Theta = 4$ orientations between zero and $\pi$. Starting from $N_\text{side} = 64$ at $j = 0$, each convolution is carried out at resolution $N_\text{side} = 64 \cdot 2^{-j}$, so that a larger $j$ corresponds to a larger angular scale on the sky. All coefficients are computed with the \texttt{FOSCAT}\footnote{\href{https://github.com/jmdelouis/FOSCAT}{https://github.com/jmdelouis/FOSCAT}.} package \cite{Delouis22, Campeti25}, which provides fast, GPU-accelerated wavelet convolutions in pixel space on the sphere.

The ST statistics provides a family of coefficients conventionally denoted $S_1$, $S_2$, $S_3$ and $S_4$. In this work we use $S_1$ and $S_2$ as our main summary statistics. For a single field $I$ they read as:
\begin{equation}
  S_1(\lambda) = \langle |I \star \psi_\lambda| \rangle \, , \qquad
  S_2(\lambda) = \langle |I \star \psi_\lambda|^2 \rangle \, ,
  \label{eq:S1S2}
\end{equation}
while for two fields $I_1$ and $I_2$, they generalise to the cross-coefficients
\begin{equation}
  S_1^\times(\lambda) = \langle |(I_1 \star \psi_\lambda)(I_2 \star \psi_\lambda)^\ast| \rangle \, , \qquad
  S_2^\times(\lambda) = \langle (I_1 \star \psi_\lambda)(I_2 \star \psi_\lambda)^\ast \rangle \, ,
  \label{eq:S1S2cross}
\end{equation}
where $\star$ is the convolution, $|\cdot|$ the modulus, and $\langle \cdot \rangle$ the spatial average. The dimension of each of these statistics is $J \cdot \Theta$. In practice, the $S_2$ coefficient characterise the auto- and cross-power spectrum of the signal on the band of scales probed by $\psi_\lambda$, while the additional information brought by $S_1$ characterises the sparsity of the signal on these bands.

\begin{figure}[t]
    \centering
    \includegraphics[width=1\textwidth]{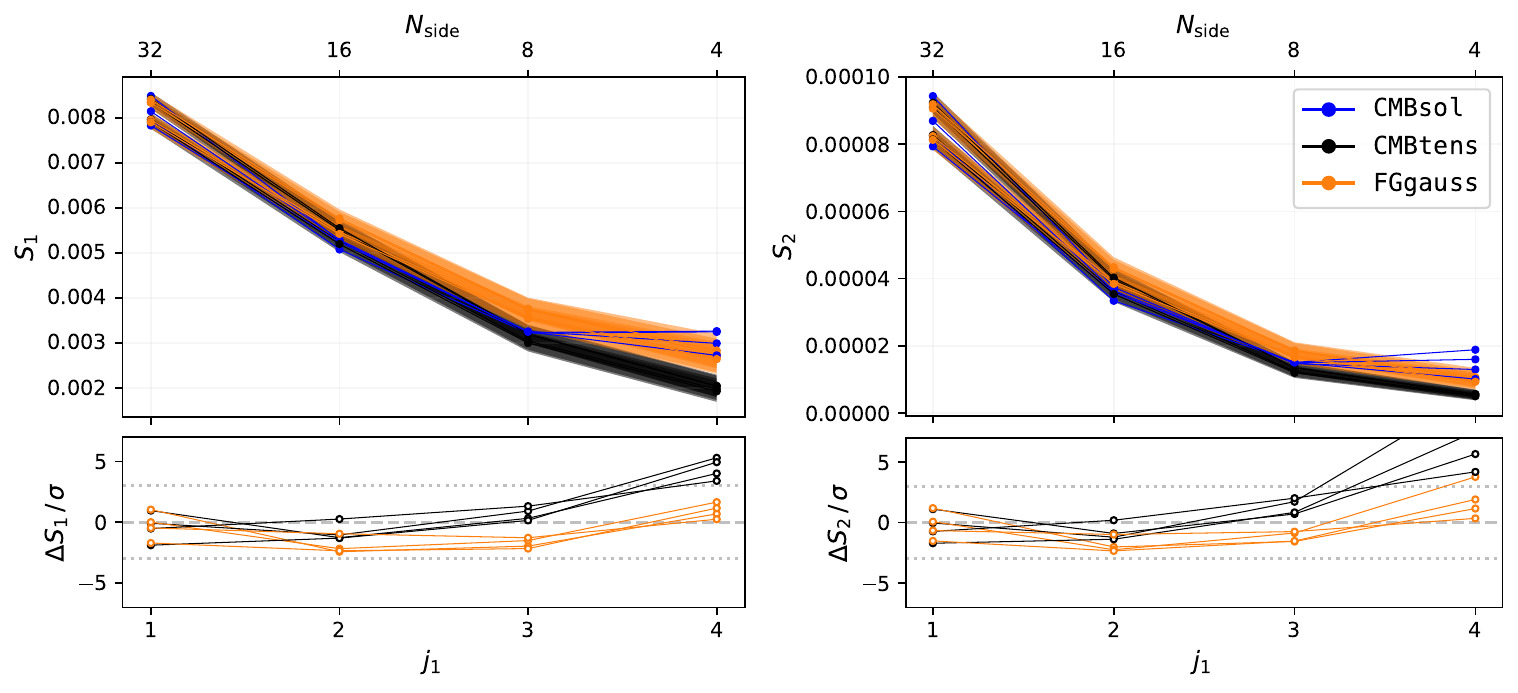}
    \includegraphics[width=1\textwidth]{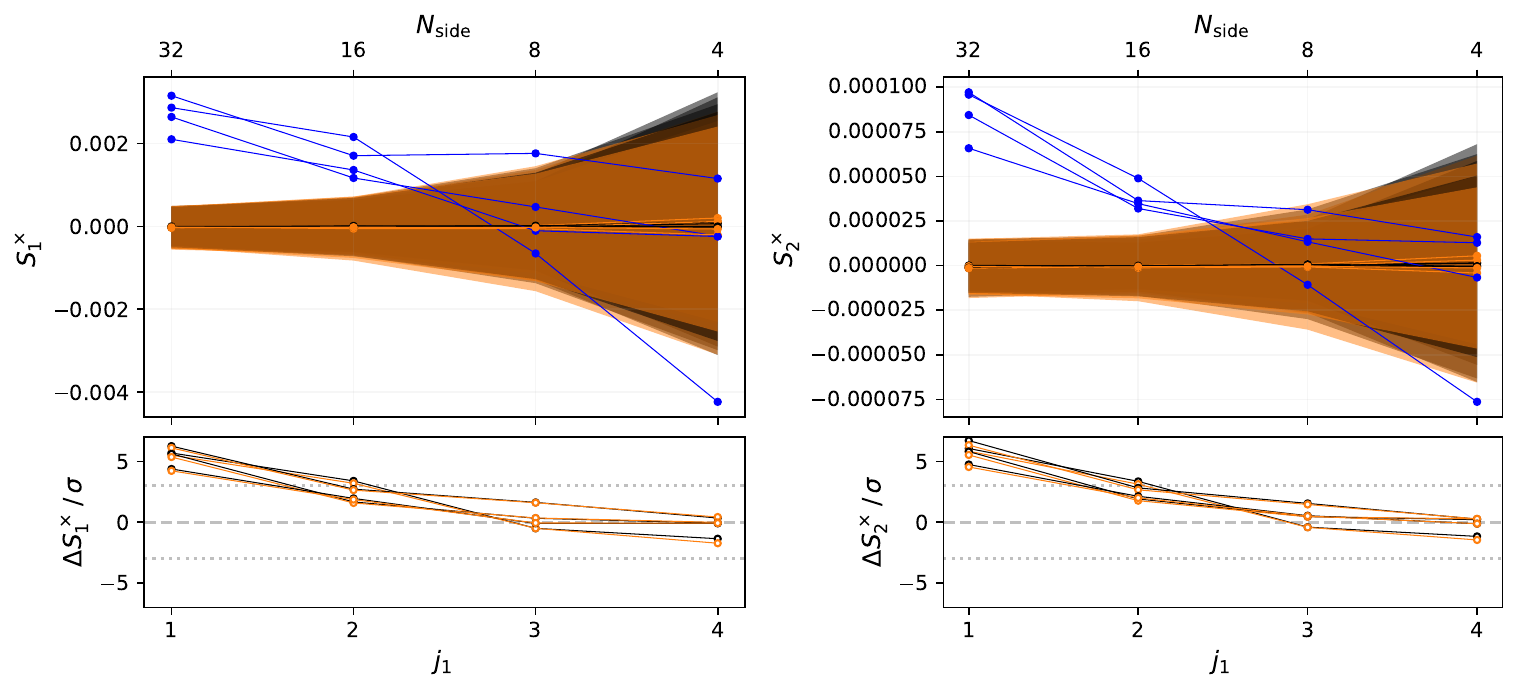}
    \caption{Scattering transform statistics, for one simulation, for the \texttt{d10s5} model with $r_\text{in} = 0$. Same-coloured lines represent different orientations. Coefficients of the recovered CMB map are in blue, while black (orange) are the average coefficients of the \texttt{CMBtens} (\texttt{FGgauss}) Gaussian simulations. The shaded areas indicate the $\pm 1 \sigma$ regions, $\sigma$ being the standard deviation of the Gaussian simulations. The secondary $x$-axis is the $N_\text{side}$ correspondent to the scale $j$. \emph{Upper figure:} ST auto-coefficients, for the NILC method. \emph{Lower figure:} ST cross-coefficients, correlating with synchrotron-dominated input data at 40 GHz, for the MC-NILC method. \emph{Bottom panels:} difference in $\sigma$ units between \texttt{CMBsol} and Gaussian maps coefficients. The dotted lines delimit the $\pm 3\sigma$ region.}
    \label{fig:s1s2}
\end{figure}

Figure \ref{fig:s1s2} shows the computed $S_1$, $S_1^\times$, $S_2$, and $S_2^\times$ coefficients for one simulation in the \texttt{d10s5} scenario, after the application of the MC-NILC component separation, with an input tensor-to-scalar ratio $r_\text{in}  = 0$. The shape of the auto-coefficients features a larger power at small scales (low $j$) where noise dominates, while it goes to zero at larger scales (high $j$). The rise of \texttt{CMBsol} coefficients at large scales is induced by the presence of non-Gaussian foreground residuals, and it is not present in the Gaussian simulations.

Each of these coefficients characterises a single oriented scale $\lambda$: it probes the structure contained within an individual wavelet band, but does not encode the coupling between distinct oriented scales. Such couplings are captured by the higher-order coefficients $S_3$ and $S_4$. Although these higher-order coefficients contain in principle a lot of additional information about non-Gaussianity, we empirically found that their greater estimator variance made them less effective than $S_1$ and $S_2$ at identifying a faint signal from a finite set of reference simulations, in our particular case. We therefore restrict our main analysis to the $S_1$ and $S_2$ coefficients. The definitions and detailed results of $S_3$ and $S_4$ are reported in appendix \ref{sec:st_appendix}.

\subsection{The robustness test}
\label{sec:test}
As mentioned in section \ref{sec:sims_gauss}, the idea of the robustness test is to compare the \texttt{CMBsol} field with reference Gaussian simulations (\texttt{CMBtens} and \texttt{FGgauss}), through the adoption of ST as the summary statistics. The employment of both the auto- and cross-coefficients allows to investigate different properties of the maps:
\begin{itemize}
    \item \textbf{non-Gaussianity.} Auto-coefficients probe non-Gaussian features associated to foreground residuals. We are mainly looking for deviations between the coefficients of the \texttt{CMBsol} map and the coefficients of \texttt{CMBtens} and \texttt{FGgauss}. We expect these deviations to be more prominent at large scales, where foreground residuals are dominant.

    \item \textbf{correlated structures.} Cross-coefficients highlight instead the presence of correlated structures between the foreground residuals in \texttt{CMBsol} (taken as $I_1$) and another selected map $I_2$ tracing the foreground emission.
\end{itemize}
For the latter, in this work we adopt the input low (40 GHz) and high (402 GHz) frequency co-added maps, described in section \ref{sec:sims}. We use these frequency channels because they are the most foreground-dominated bands available in our experimental setup, with synchrotron and dust dominating the emission at 40 GHz and 402 GHz, respectively. If the foreground residuals in \texttt{CMBsol} induce a bias on $r$, we expect to find some correlation with structures in foregrounds-dominated maps, which are instead absent in pure Gaussian simulations. Consequently, we expect to see incompatible ST cross-coefficients.

Ideally, one would like to correlate with foregrounds-only sky maps, where dust or synchrotron are the only emissions present. In general, this is not possible for a real experiment, even if component separation methods able to remove part of the CMB and noise contributions are available \cite{Planck20_compsep}. An alternative is to use external tracers of foreground emission as the correlation target, for instance \emph{WMAP} \cite{WMAP13_results} or \emph{Planck} \cite{Planck20_dust} low/high frequency maps, even if these datasets are expected to have a larger noise with respect to the aforementioned internal maps. We choose to not use external data here, since our foregrounds simulations from \texttt{PySM} heavily rely on the information contained in \emph{Planck} and \emph{WMAP} observations, with the risk of invalidating the analysis. Instead, it is always possible to use the multi-frequency sky maps observed by the same experiment as a correlating field, even if they have some ``contamination" due to CMB and noise. In more detail, by construction of the component separation algorithm, the CMB and noise residuals contained in \texttt{CMBsol} will be correlated with the original CMB and noise present in the data, since the component separation is just a linear combination of the input maps. The noise correlation will be more evident for the lowest and highest frequencies (40 and 402 GHz), i.e. the noisiest channels. These spurious correlations are instead absent for Gaussian simulations. Therefore, the cross-coefficients will register some failures of the robustness test (false positives) not due to foregrounds contamination, but induced by CMB and noise correlations. The impact of these false positives is quantified in the validation phase of the test, reported in the next section.

We summarise here the main steps of the robustness test:
\begin{enumerate}
    \item we use the tensor-to-scalar ratio $r_\text{fgs}$ measured from the component-separated \texttt{CMBsol} to generate 5000 Gaussian simulations (\texttt{CMBtens}) for each of the 300 \texttt{CMBsol} realisations. We also replace the foreground residuals component with a Gaussian version of it, generating another set of simulations (\texttt{FGgauss});

    \item we evaluate the auto- and cross-ST coefficients $S$ and $S^\times$ of \texttt{CMBsol} and \texttt{CMBtens} (and \texttt{FGgauss}), and compare them through a compatibility analysis (described in appendix \ref{sec:compatibility}):
    \begin{equation}
        \begin{aligned}
        S \left( \texttt{CMBsol} \right) \quad & \text{vs} \quad S \left( \texttt{CMBtens} \right) \\
        S^\times \left( \texttt{CMBsol} \times 40 \text{ GHz} \right) \quad & \text{vs} \quad S^\times \left( \texttt{CMBtens} \times 40 \text{ GHz} \right) \\
        S^\times \left( \texttt{CMBsol} \times 402 \text{ GHz} \right) \quad & \text{vs} \quad S^\times \left( \texttt{CMBtens} \times 402 \text{ GHz} \right),
        \end{aligned}
    \label{eq:coeffs_test}
    \end{equation}
    where $S$ stands either for $S_1$ or $S_2$;

    \item if the coefficients are compatible, the test is passed and no bias is identified. If the coefficients are not compatible, the test is failed and the bias is identified. We report the efficiency $\eta$ of the robustness test as the percentage of simulations where a bias has been identified, for the different scenarios considered:
    \begin{equation}
        \eta = 100 \times \frac{\text{number of failing sims}}{\text{number of sims}} \, .
    \end{equation}
    A value of $\eta$ close to 100\% suggests that the test can confidently recognise a bias on $r$, while a lower value indicates that the test is not sensitive enough to be reliable.
\end{enumerate}

\section{Results and discussion}
\label{sec:results}
In this section we present and discuss the results of the characterisation of the ST-based robustness test. Figure \ref{fig:s1s2} shows an example of the $S_1$ and $S_2$ coefficients computed on the three types of maps described in the previous sections, after the application of MC-NILC component separation on the \texttt{d10s5} foreground model, with no primordial signal ($r_\text{in} = 0$). Table \ref{tab:eta_j14} reports the efficiency of the test $\eta$ for the various considered scenarios. The ``auto" columns list results for the auto-coefficients, while the $\times$ sign indicates cross-correlation with foregrounds-dominated frequency maps. For the purposes of the compatibility analysis, the two coefficients are combined together to form a single array, $S_{12} = S_1 \cup S_2$. A scenario where the CMB lensing contribution is reduced by a factor $A_\text{lens} = 0.5$ is also considered, which is a typical delensing value forecasted to be achieved by current and future experiments \cite{SO19, Hertig24, SO25}.

\begin{table}
    \caption{Efficiency $\eta$ (in \%) of the ST-based robustness test on polarisation $B$-mode maps, for the different scenarios considered in the study. Here we are exploiting the $S_1$ and $S_2$ coefficients combined together into a single array, between scales $j_\text{min} = 1$ and $j_\text{max} = 4$. This range of scales is selected to avoid the noise correlation described in the text.}
    \label{tab:eta_j14}
    \begin{center}
        \begin{tabular}{l c c c c c}
            \multicolumn{6}{c}{MC-NILC} \\
            \hline
            & reference & auto & auto & $\times$ 402 & $\times$ 40 \\
            & sims & $A_\text{lens} = 1$ & $A_\text{lens} = 0.5$ & GHz & GHz \\
            \hline
            \multirow{2}{1.5cm}{\texttt{d0s0}\\$r_\text{in} = 0$} & \texttt{CMBtens} & 5 & 7 & 6 & 8 \\
            & \texttt{FGgauss} & 5 & 6 & 6 & 8 \\
            \hline
            \multirow{2}{1.5cm}{\texttt{no fgs}\\$r_\text{in} > 0$} & \texttt{CMBtens} & 6 & $\circ$ & 6 & 16 \\
            & \texttt{FGgauss} & 6 & $\circ$ & 5 & 15 \\
            \hline
            \multirow{2}{1.5cm}{\texttt{d10s5}\\$r_\text{in} = 0$} & \texttt{CMBtens} & 52 & 66 & 45 & 93 \\
            & \texttt{FGgauss} & 29 & 45 & 50 & 92 \\
            \hline
            \\
            \multicolumn{6}{c}{NILC} \\
            \hline
            \multirow{2}{1.5cm}{\texttt{d0s0}\\$r_\text{in} = 0$} & \texttt{CMBtens} & 4 & 6 & 5 & 13 \\
            & \texttt{FGgauss} & 4 & 6 & 5 & 13 \\
            \hline
            \multirow{2}{1.5cm}{\texttt{d10s5}\\$r_\text{in} = 0$} & \texttt{CMBtens} & 100 & 100 & 96 & 98 \\
            & \texttt{FGgauss} & 88 & 93 & 81 & 90 \\
            \hline
        \end{tabular}
    \end{center}
\end{table}

\paragraph{MC-NILC - \texorpdfstring{\texttt{d10s5} - $r_\text{in} = 0$}{d10s5 - r = 0}.} This is the baseline scenario for the study, as it consider the most advanced component separation method applied to a realistic foregrounds model. The bias on the tensor-to-scalar ratio\footnote{The obtained bias here is also related to the sub-optimal mask we are using, since it only covers the Galactic plane. In a real experiment, a more optimised mask would be adopted and the bias would be lower. Scenarios with lower foreground residuals and smaller biases are investigated later in the section.} is about $\delta(r) \sim 3 \times 10^{-3}$. Using auto-coefficients, the ST test identifies a bias on $r$ in 52\% and 29\% of the simulations when \texttt{CMBsol} is compared with the \texttt{CMBtens} and \texttt{FGgauss} simulations, respectively. The efficiency improves to 66\% and 45\% when we reduce some of the CMB lensing Gaussian contribution, sharpening the foreground residuals features. We notice the lower efficiency obtained when comparing with \texttt{FGgauss} simulations, which are more similar to \texttt{CMBsol} maps in terms of power spectrum. Thus, for this particular scenario, auto-coefficients are not reliable enough to consistently identify the bias on $r$, as the test only raises a flag in roughly half of the simulations. The same is true if we consider the cross-coefficients computed by correlating with the dust-dominated data at 402 GHz, with the efficiency being around 50\%. However, the test is very efficient when cross-correlating with synchrotron-dominated data at 40 GHz, being $\eta \sim 92\%$ both when comparing with \texttt{CMBtens} and \texttt{FGgauss}. This means that in case of a $\delta(r) \sim 10^{-3}$ bias, the robustness test identifies it in almost the totality of the simulations, raising a flag on the presence of synchrotron-correlated foreground residuals.

\paragraph{NILC - \texorpdfstring{\texttt{d10s5} - $r_\text{in} = 0$}{d10s5 - r = 0}.} Compared with MC-NILC, the NILC technique leaves a higher amplitude of foregrounds contamination at large scales, leading to a higher bias of $\delta(r) \gtrsim 5 \times 10^{-3}$. In this case, both the auto- and cross-coefficients are able to identify the bias in a large number of simulations, with the efficiency reaching $\sim 90\%$ and even 100\% in the \texttt{FGgauss} and \texttt{CMBtens} comparisons, respectively.

\paragraph{Validation.} The validation of the robustness test is a crucial step for its characterisation. We consider three different validation scenarios:
\begin{itemize}
    \item we adopt the simpler foregrounds model \texttt{d0s0}. Component separation can easily mitigate this contamination and the input tensor-to-scalar ratio $r_\text{in} = 0$ is recovered with no or very small bias, $\delta(r) \lesssim 10^{-4}$. In this case the efficiency of the test is around $\sim 5-10\%$ for both MC-NILC and NILC, for all scenarios, with the small number of incompatibilities due to chance or noise correlations in the samples. Thus, if $r$ is not biased, the robustness test is passed and no flag is raised.

    \item We consider a scenario where, in addition to the $B$-modes induced by gravitational lensing, a realistic primordial signal is also included in the input simulations, with a tensor-to-scalar ratio of $r_\text{in} = 5 \times 10^{-3}$. To recover an unbiased estimate of $r$, we manually remove the foreground residuals from the component separated CMB maps\footnote{We recall that, since we are working with simulated data, we have all the component separated maps available.}. The efficiency is around $\sim 5-10\%$ as in the previous case, certifying that the incompatibilities measured in the \texttt{d10s5} scenario are entirely due to the presence of foreground residuals, and not induced by other components. The slightly higher number of false positives in the cross-40 GHz coefficients (15\%) quantifies the impact of the CMB and noise residuals correlation with the input data (see the discussion in section \ref{sec:test}). Compared to the previous \texttt{d0s0} - $r_\text{in} = 0$ case, there are more false positives because now there is an additional primordial signal in the CMB, strengthening the correlation. The contribution due to noise correlation is detailed in appendix \ref{sec:noise_corr}.

    \item To further confirm that the test is really flagging the presence of foregrounds, we repeat the analysis in the MC-NILC - \texttt{d10s5} - $r_\text{in} = 0$ scenario, computing cross-coefficients with the pure-synchrotron maps at 40 GHz to keep the foregrounds-only correlation and avoid CMB and noise correlations. We apply a rescaling factor $A_\text{fgs} < 1$ of different values to the foreground residuals contained in \texttt{CMBsol} (lower left map of figure \ref{fig:maps_test}). In this way, we gradually lower the foreground residuals amplitude and the $r$ bias. If the incompatibilities detected by the test are only due to foreground residuals, then the efficiency should decrease and approach the $\sim 5-10\%$ range. This step also provides the test sensitivity as a function of the bias on $r$: for each decreasing value of $A_\text{fgs}$, the bias becomes lower, and we measure in how many simulations the bias is identified. The left panel of figure \ref{fig:pte_vs_bias_cross_cl} shows the PTE ($p$-value) given by the compatibility analysis described in appendix \ref{sec:compatibility} for different values of the $r$ bias, for 50 simulations. If the PTE for a simulation is less than the 0.05 threshold (dashed horizontal line in the figure), contamination has been recognised by the test. From the plot, when the bias is $\delta(r) \gtrsim 1 \times 10^{-3}$, the robustness test can confidently identify it, with PTEs being $< 0.05$ for almost all simulations. When the bias is lower than $\sim 5 \times 10^{-4}$ the test is not able to confidently flag the bias any more, as the PTEs increasingly exceed the 0.05 threshold. This immediately shows that foreground residuals are responsible for the incompatibilities previously reported, as it is the only component we are reducing here. In the context of a \emph{LiteBIRD}-like experiment, these results are additionally promising because $\simeq 10^{-3}$ is the target sensitivity of future missions searching for primordial $B$-modes. Any bias lower than this approximate threshold will be confused with noise and thus it will not induce a false $r$ detection. In any case, the value of the lowest identifiable bias will be dependent on the explored instrumental configuration, and in particular on the achievable noise level.
\end{itemize}
Thus, the validation stage consolidates the reliability of the ST-based robustness test. Results show how the test can confidently warn about the presence of foreground residuals contamination in a biased detection of the tensor-to-scalar ratio, while it raises no strong flags when the estimate is unbiased.

\begin{figure}[t]
    \centering
    \includegraphics[width=1.\textwidth]{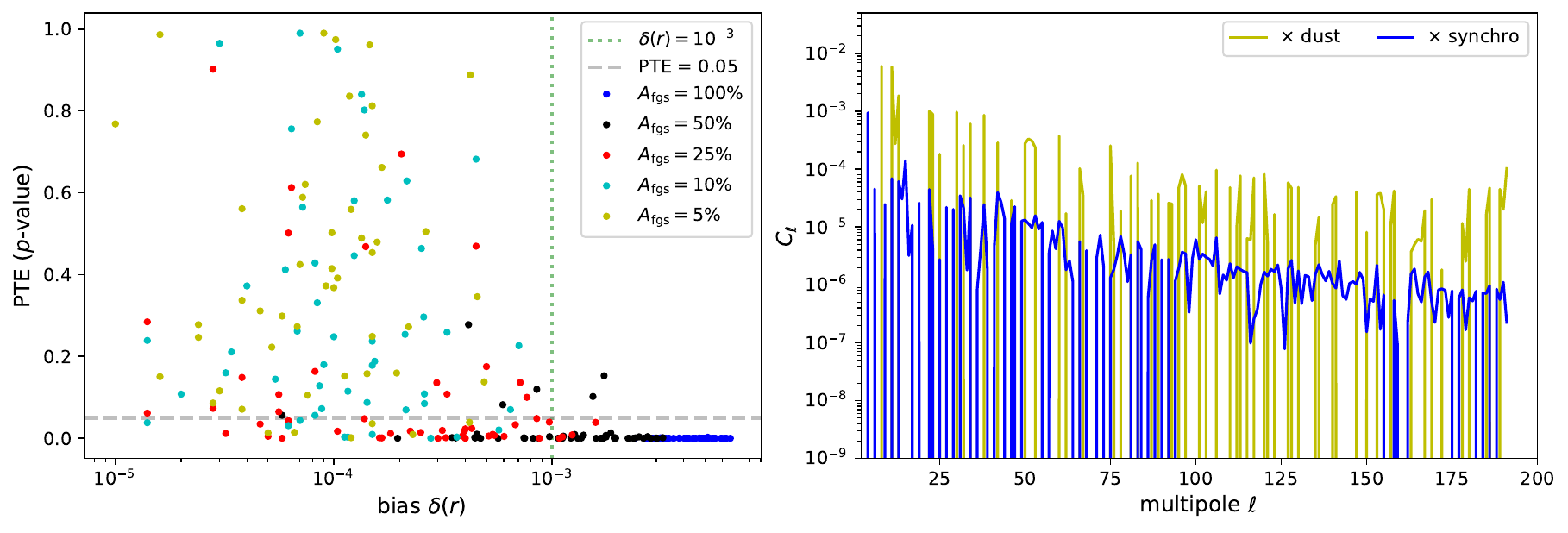}
    \caption{The considered scenario is MC-NILC - \texttt{d10s5} - $r_\text{in} = 0$. \emph{Left panel:} probability-to-exceed (PTE, or $p$-value) given by the compatibility analysis for the cross-synchrotron test, as a function of the bias on the tensor-to-scalar ratio $r$. Results are shown for 50 simulations and for different values of foregrounds rescaling $A_\text{fgs}$. The dotted vertical line is a reference value of $\delta(r) = 10^{-3}$, while the horizontal dashed line is the PTE threshold for incompatibility: above 0.05, coefficients are compatible, under 0.05, they are not, and foreground contamination is detected. \emph{Right panel:} cross-angular power spectrum between \texttt{CMBsol} and foreground maps; cross-correlation with dust is indicated in yellow, while blue is for synchrotron. See text for discussion.}
    \label{fig:pte_vs_bias_cross_cl}
\end{figure}

\paragraph{Comparison with other statistics.} For completeness, we also compare the above performances with other summary statistics that could be used for the same robustness test. Table \ref{tab:eta_cl_j14} reports the test efficiency when the usual isotropic angular power spectrum $C_\ell$ is adopted instead of ST coefficients. In this case we can only rely on the cross-data power spectra, since the Gaussian simulations are built to have auto-spectra similar to \texttt{CMBsol}. Compared to ST coefficients in the baseline scenario, the isotropic power spectrum performs way worse, with the percentage of identified biased simulations being $\sim 30\%$ and 10\% against the previous $\sim 90\%$ and 50\%, when correlating with synchrotron- and dust-dominated frequency maps, respectively. This clearly shows the advantage of the ST approach: with respect to the simpler isotropic power spectrum, ST coefficients retain precious information about the oriented scales through the angular dependency of the wavelet filter. \\
We can also compare ST statistics with the results obtained in the previous part of this work, where Minkowski Functionals (MFs) have been employed for the same scope (tables 2 and 3 of \cite{Ranucci26}). MFs can detect non-Gaussian foreground residuals in just 27\% and 13\% of the simulations when comparing with \texttt{FGgauss} maps, for NILC and MC-NILC respectively, applied on the \texttt{d10s5} model. In the same scenario, they are clearly outperformed by ST statistics, which obtain instead 88\% and 29\% using auto-coefficients, and $\sim 90\%$ when adopting cross-coefficients correlating with low frequency data. Again, the combination of multi-scale approach, angular orientation, and low variance allows the ST coefficients to stand out in the search for non-Gaussian features, when compared against other commonly used summary statistics.

\begin{table}
    \caption{Efficiency $\eta$ (in \%) of the $C_\ell$-based test on polarisation $B$-mode maps, for the different scenarios considered in the study. Here we are exploiting scales between $\ell_\text{min}=11$ and $\ell_\text{max}=95$ (similar to the scales sampled by $j \in [1,4]$ in table \ref{tab:eta_j14}) to avoid noise correlation. Results are shown for the MC-NILC component separation method.}
    \label{tab:eta_cl_j14}
    \begin{center}
        \begin{tabular}{l c c c}
            \hline
            & reference & $\times$ 402 & $\times$ 40 \\
            & sims & GHz & GHz \\
            \hline
            \multirow{2}{1.5cm}{\texttt{d0s0}\\$r_\text{in} = 0$} & \texttt{CMBtens} & 2 & 12 \\
            & \texttt{FGgauss} & 2 & 12 \\
            \hline
            \multirow{2}{1.5cm}{\texttt{no fgs}\\$r_\text{in} > 0$} & \texttt{CMBtens} & 4 & 15 \\
            & \texttt{FGgauss} & 2 & 12 \\
            \hline
            \multirow{2}{1.5cm}{\texttt{d10s5}\\$r_\text{in} = 0$} & \texttt{CMBtens} & 10 & 34 \\
            & \texttt{FGgauss} & 10 & 27 \\
            \hline
        \end{tabular}
    \end{center}
\end{table}

The last point to clarify concerns the higher test efficiency we get by correlating with low frequency maps, in comparison with high frequency, both for ST coefficients and power spectra. This is not obvious, as we generally expect dust to have a similar or larger impact than synchrotron on the $r$ recovery. An immediate way to understand this is to look at the cross angular power spectra between \texttt{CMBsol} and the raw input foregrounds maps at 40 and 402 GHz. Power spectra are shown in the right panel of figure \ref{fig:pte_vs_bias_cross_cl}. Correlation with dust maps is compatible with zero, with values oscillating between positive and negative. Instead, we see a clear positive correlation with synchrotron maps, at intermediate scales (from $\ell \sim 90$), which are the same scales where we observe the main difference between \texttt{CMBsol} and Gaussian simulations in the ST coefficients ($j = 1$ in figure \ref{fig:s1s2}). This correlation suggests the presence of synchrotron-related structures at intermediate scales in the foreground residuals, which seems to be less mitigated by the foreground-cleaning algorithms adopted in this work. This is also a consequence of the adopted instrumental configuration; synchrotron radiation is monitored through the lowest frequency bands, which in our case is the (not-so-low) 40 GHz channel, while thermal dust is characterised in several high-frequency bands (280, 337, 402 GHz). We notice here that different component separation methodologies, not based on the internal linear combination approach, may lead to different properties of the foreground residuals maps. This might consequently result in different test efficiencies. It is worth stressing that the robustness test methodology presented in this work is completely general and can be straightforwardly applied to other map-based component separation techniques (parametric, template-fitting, etc.), keeping in mind that each method requires proper characterisation and validation stages similar to the ones described here.

\section{Conclusions}
\label{sec:conclusions}
In this work, we proposed and presented the performances of a robustness test for the validation of a potential detection of the tensor-to-scalar ratio from CMB observations, using scattering transforms (ST) to identify the bias induced by the presence of leftover foreground residuals in component-separated CMB maps. We characterised and validated the robustness test on the instrumental configuration of a \emph{LiteBIRD}-like experiment, applying it to different sets of simulations with simple (\texttt{d0s0}) and more complex (\texttt{d10s5}) models of foreground emission, adopting blind component separation algorithms.

The goal of the proposed robustness test is to reject a potential detection of primordial $B$-modes by looking for foregrounds-related non-Gaussian features in the CMB field recovered from multi-frequency sky maps through a map-based component separation. The ST coefficients of the contaminated CMB map are compared with the coefficients of Gaussian simulations of primordial $B$-modes: if the statistics are compatible, the test is passed; if the statistics are not compatible, the test is failed and a warning is raised. 

We demonstrated that the ST-based robustness test can confidently flag the presence of a bias $\delta(r)$ in almost all of our simulations, when $\delta(r) \gtrsim 10^{-3}$, a threshold similar to the sensitivity targeted by \emph{LiteBIRD} for the measurement of $r$. We investigated both auto- and cross-ST coefficients, in order to exploit the non-Gaussian properties of foreground residuals and the correlation of structures with foregrounds-dominated frequency maps. As validation, we also showed that the test raises no strong warning in case of an unbiased detection of $r$, or in the absence of foreground contamination.

We also compared the ST performances with respect to other commonly used statistics, as the isotropic angular power spectrum and Minkowski functionals \cite{Ranucci26}. Results showed how ST coefficients outperform the other statistics in every considered scenario, with a clear improvement in recognising foreground residuals in component-separated CMB maps. The combination of a low-variance multi-scale approach, and estimation of statistical dependency that go beyond linear correlation (with modulus) makes ST statistics extremely sensitive to the features contaminating a detection of the tensor-to-scalar ratio.

It is worth noticing that recent works \cite{Tsouros26} suggest the possibility for the \texttt{PySM} models (as the \texttt{d10} used here) to not be able to properly reproduce the sharpest non-Gaussian features present in the real sky. This implies that, when applied on real data, the proposed methodology could face a statistically more non-Gaussian sky than the considered simulations, with structures being possibly easier to identify. The test could thus perform differently in a real experiment than what we have presented, with the possibility to get to a higher efficiency.

The methodology described in this work can be applied to any map-based component separation method (parametric, blind, template-fitting). It is also completely general, as it can be applied to a variety of scenarios, including different experimental configurations (\emph{LiteBIRD}, \emph{SO}) or different contaminations (foregrounds, systematics). The only crucial point is the need to completely characterise the robustness test in selected scenarios, as we have shown in this paper, in order to fully explore the test capabilities and establish a benchmark for a fair comparison with a future real measurement of $r$.

We conclude by remarking the importance of being prepared for the results coming from current and future experiments targeting the measurement of primordial $B$-modes. As the sensitivity of CMB observatories improves every year, a potential detection is rapidly becoming a possible outcome. A complete battery of robustness tests is fundamental for the validation, acceptance, or rejection of an measurement of the tensor-to-scalar ratio, in order for the community to avoid misleading claims. This work provides an important step forward in this direction.

\acknowledgments
We thank Matthieu Tristram, Sophie Henrot-Versill\'e, Thibaut Louis, Alexandros Tsouros, Cl\'ement Leloup, Florie Carralot, and BB for the many useful comments and discussions during the preparation of the paper. CR also thanks the entire CMB group at IJCLab for their kind hospitality and support during the development of the project. Authors from SISSA acknowledge partial support by the Italian Space Agency (ASI Grants No.~2020-9-HH.0 and 2016-24-H.1-2018) and the Euclid Project, as well as the RadioForegroundsPlus Project HORIZON-CL4-2023-SPACE-01, GA 101135036 and through the Project SPACE-IT-UP by the Italian Space Agency and Ministry of University and Research, Contract Number 2024-5-E.0. They also acknowledge partial support by the InDark and \emph{LiteBIRD} Initiative of the National Institute for Nuclear Physics. EA and SP received government funding managed by the French National Research Agency under France 2030, reference number ``ANR-25-CE46-6634", as well as from the Paris Region under the DIM ORIGINE funding. PC is funded by the European Union (ERC, RELiCS, project number 101116027). Views and opinions expressed are however those of the authors only and do not necessarily reflect those of the European Union or the European Research Council Executive Agency. Neither the European Union nor the granting authority can be held responsible for them.

In this work we made use of the following additional software/packages: \href{https://numpy.org/}{\texttt{numpy}} \cite{numpy}, \href{https://scipy.org/}{\texttt{scipy}} \cite{scipy}, \href{https://www.astropy.org/}{\texttt{astropy}} \cite{astropy}, \href{https://matplotlib.org/}{\texttt{matplotlib}} \cite{matplotlib}. Part of this research used resources of the National Energy Research Scientific Computing Center (\href{https://www.nersc.gov/}{NERSC}), a Department of Energy User Facility (projects \texttt{mp107b-2025/2026} and \texttt{mp107d-2025/2026}).

\appendix

\section{Component separation}
\label{sec:compsep}
In this section we report some additional details on the component separation techniques we adopted in this work. We still point the interested reader to \cite{Carones23_nilc, Carones23_mcnilc, Carones26_broom} for complete descriptions and discussions of the algorithms.

NILC \cite{Delabrouille09_nilc} is a component separation method that aims at reconstructing the CMB signal through a minimum-variance approach, in order to reduce foreground contamination without any assumptions on the foreground spectral properties. It thus represents a valuable alternative to parametric approaches, as it is not affected by spectral mis-modelling of the Galactic polarised emission, which may significantly bias the final estimate of the tensor-to-scalar ratio. It performs Internal Linear Combination (ILC) using a spherical wavelet (or needlet) analysis of the multi-frequency sky maps. This enable localised processing both in pixel and needlet domain, and full exploitation of large scale correlations of the CMB and foreground emissions.

In general, foreground emission show strong variations across the sky, with foreground properties being very different between the higher latitude and near the Galactic plane. Since NILC performs simple variance minimisation across the entire sky, it does not fully handle the foregrounds local spectral variations. MC-NILC \cite{Carones23_mcnilc} is a NILC extension which performs component separation independently in different regions of the sky, where the polarised $B$-mode Galactic emission shows similar spectral properties. To do so, the spatial variability of dust and synchrotron spectral parameters is assessed by identifying a blind tracer of their distribution across the sky, generally estimated through the ratio of different frequency maps.

Both algorithms relies on the choice of the needlet system. In this work, we adopted a set of needlet filters with a ``mexican" harmonic function and a bandwidth parameter of 1.3. The first 14 bands are merged together in order to include more modes in the first band. Needlet bands are computed with the \texttt{MTNeedlet}\footnote{\href{https://javicarron.github.io/mtneedlet/index.html}{https://javicarron.github.io/mtneedlet/index.html}} package \cite{CarronDuque19}.

\section{Additional ST statistics}
\label{sec:st_appendix}
\begin{figure}
    \centering
    \includegraphics[width=1\textwidth]{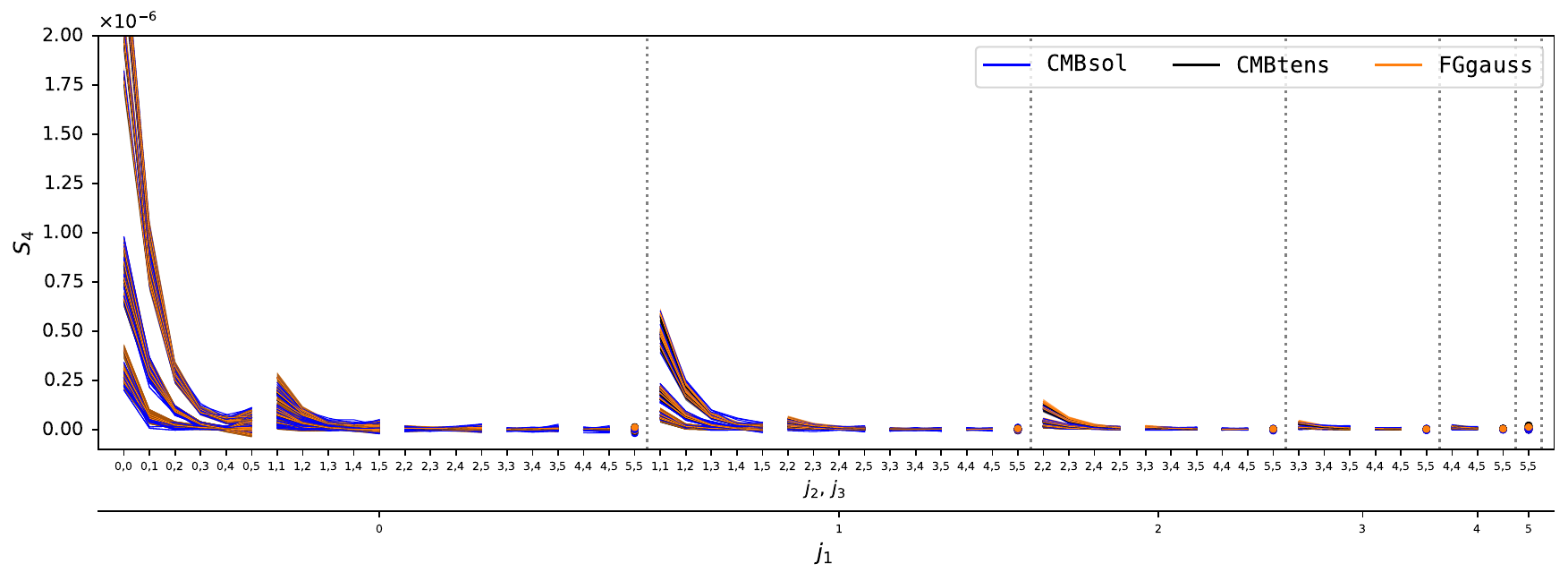}
    \includegraphics[width=1\textwidth]{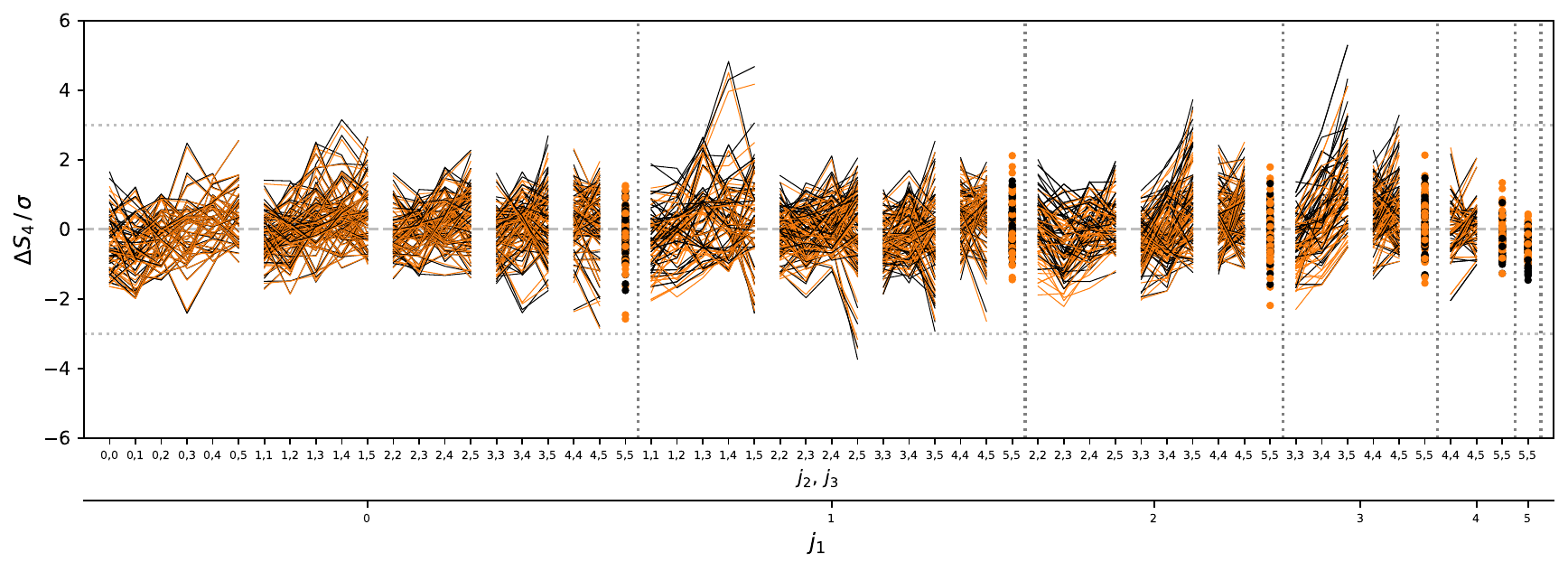}
    \caption{$S_4$ auto-coefficient for the MC-NILC - \texttt{d10s5} - $r_\text{in} = 0$ scenario. Colour code is the same as figure \ref{fig:s1s2}. The lower plot shows the residual difference between \texttt{CMBsol} and Gaussian simulations, in units of the standard deviation of the simulations $\sigma$.}
    \label{fig:s4}
\end{figure}

The $S_1$ and $S_2$ coefficients introduced in section \ref{sec:st} each describe a single oriented scale. Within the scattering covariance formalism \cite{Morel22, Cheng24, Mousset24}, the coupling between scales is captured by the $S_{3}$ and $S_{4}$ coefficients. They are defined for a single field $I$ as:
\begin{equation}
    \begin{aligned}
        S_3(\lambda_1, \lambda_2) & = \mathrm{Cov} \left[ I \star \psi^{\lambda_1}, \, | I \star \psi^{\lambda_2} | \star \psi^{\lambda_1} \right], \\
        S_4(\lambda_1, \lambda_2, \lambda_3) & = \mathrm{Cov} \left[| I \star \psi^{\lambda_3} | \star \psi^{\lambda_1}, \, | I \star \psi^{\lambda_2} | \star \psi^{\lambda_1} \right],
    \end{aligned}
    \label{eq:coeffs_s4}
\end{equation}
where $\lambda_i = (j_i, \theta_i)$ specifies the scale $j_i$ and orientation $\theta_i$ of each wavelet filter, and the covariance between two complex fields $X$ and $Y$ is defined as $\mathrm{Cov}[X, Y] = \langle XY^* \rangle - \langle X \rangle \langle Y^* \rangle$. As for the $S_{1}$ and $S_{2}$ coefficients, they extend to cross-statistics between two fields $I_1$ and $I_2$:
\begin{equation}
    \begin{aligned}
        S_3^\times (\lambda_1, \lambda_2) & = \mathrm{Cov} \left[ I_1 \star \psi^{\lambda_1}, \, | I_2 \star \psi^{\lambda_2} | \star \psi^{\lambda_1} \right], \\
        S_4^\times (\lambda_1, \lambda_2, \lambda_3) & = \mathrm{Cov} \left[| I_1 \star \psi^{\lambda_3} | \star \psi^{\lambda_1}, \, | I_2 \star \psi^{\lambda_2} | \star \psi^{\lambda_1} \right].
    \end{aligned}
    \label{eq:coeffs_s4_cross}
\end{equation}
Figure \ref{fig:s4} shows the auto-$S_4$ coefficients for the maps used in the robustness test, where the noise contribution is evident at small scales (low $j$). Table \ref{tab:eta_s4} reports the corresponding test efficiencies, obtained through the dimensionality reduction and compatibility analysis described in appendix \ref{sec:compatibility}. Since $S_3$ and $S_4$ characterise the interaction between two and three oriented scales, they are expected to give additional information than $S_1$ and $S_2$ to non-Gaussian foreground residuals \cite{Cheng24, Mousset24, Tsouros26}. In practice, however, we find that this is not the case in our specific setting. We measure a similar or slightly higher sensitivity only for the auto-coefficients for \texttt{FGgauss}, while the efficiency drops when cross-correlating with foreground-dominated maps. This is once again a consequence of estimator variance: the larger variance of $S_3$ and $S_4$ makes it difficult to separate fields that are overall dominated by Gaussian components such as CMB lensing and noise. Despite being simpler, $S_1$ and $S_2$ were found to be more powerful for the robustness test, which is why we adopt them in the main analysis.

\begin{table}
    \caption{Efficiency $\eta$ (in \%) of the ST-based robustness test on polarisation $B$-mode maps, for the different scenarios considered in the study. Here we are exploiting the $S_4$ coefficient, after the dimensionality reduction described in appendix \ref{sec:compatibility}.}
    \label{tab:eta_s4}
    \begin{center}
        \begin{tabular}{l c c c c c}
            \multicolumn{6}{c}{MC-NILC} \\
            \hline
            & reference & auto & auto & $\times$ 402 & $\times$ 40 \\
            & sims & $A_\text{lens} = 1$ & $A_\text{lens} = 0.5$ & GHz & GHz \\
            \hline
            \multirow{2}{1.5cm}{\texttt{d0s0}\\$r_\text{in} = 0$} & \texttt{CMBtens} & 5 & 6 & 1 & 2 \\
            & \texttt{FGgauss} & 4 & 6 & 1 & 2 \\
            \hline
            \multirow{2}{1.5cm}{\texttt{no fgs}\\$r_\text{in} > 0$} & \texttt{CMBtens} & 5 & $\circ$ & 4 & 4 \\
            & \texttt{FGgauss} & 3 & $\circ$ & 1 & 3 \\
            \hline
            \multirow{2}{1.5cm}{\texttt{d10s5}\\$r_\text{in} = 0$} & \texttt{CMBtens} & 42 & 52 & 38 & 38 \\
            & \texttt{FGgauss} & 39 & 56 & 33 & 35 \\
            \hline
            \\
            \multicolumn{6}{c}{NILC} \\
            \hline
            \multirow{2}{1.5cm}{\texttt{d0s0}\\$r_\text{in} = 0$} & \texttt{CMBtens} & 4 & 4 & 1 & 3 \\
            & \texttt{FGgauss} & 3 & 4 & 1 & 3 \\
            \hline
            \multirow{2}{1.5cm}{\texttt{d10s5}\\$r_\text{in} = 0$} & \texttt{CMBtens} & 100 & 100 & 78 & 75 \\
            & \texttt{FGgauss} & 89 & 93 & 51 & 48 \\
            \hline
        \end{tabular}
    \end{center}
\end{table}

\section{Compatibility analysis}
\label{sec:compatibility}
\paragraph{Dimensionality reduction} The compatibility test described below relies on the Mahalanobis distance between the ST statistics of \texttt{CMBsol} and an ensemble of ST statistics computed on reference Gaussian maps, either \texttt{CMBtens} or \texttt{FGgauss}. Evaluating this distance requires inverting the covariance $\hat{\boldsymbol \Sigma}$ of the ST statistics of the ensemble. Estimating and inverting a $D \times D$ covariance from a finite number $N$ of realisations is reliable only when $N \gg D$: as $D$ approaches $N$, the estimate becomes ill-conditioned, and its inverse strongly biased. The raw ST statistics, especially when $S_3$ and $S_4$ are included, have a dimension $D$ that is too large relative to the $N = 5000$ reference Gaussian maps available. Moreover, because the Mahalanobis distance aggregates all retained coefficients into a single scalar, coefficients that carry little information only inflate the variance of the distance, diluting the test. We therefore reduce the dimensionality of the statistics with two objectives: first, to obtain stable, well-conditioned estimates of $\hat{\boldsymbol \Sigma}$; and second, to retain the coefficients most likely to carry information about the departure from Gaussianity of \texttt{CMBsol}. Each ST coefficient depends on one (for $S_1$ and $S_2$) or several (for $S_3$ and $S_4$) scales $j$, as well as on angles $\theta$. This dependence on scale and angle is typically smooth, and we exploit this regularity to compress the representation. We Fourier transform the coefficients over the angles and keep only the zeroth mode and the first harmonic. For $S_3$ and $S_4$, we additionally retain only those coefficients whose scales lie close together: a coefficient depending on two scales $j_1$ and $j_2$ is discarded whenever $|j_1 - j_2| > d_j$, with the threshold set to $d_j = 4$. As an example, the number of values entering the compatibility analysis goes from 3584 to 512 for the $S_4$ coefficient.

\begin{figure}
    \centering
    \includegraphics[width=0.9\textwidth]{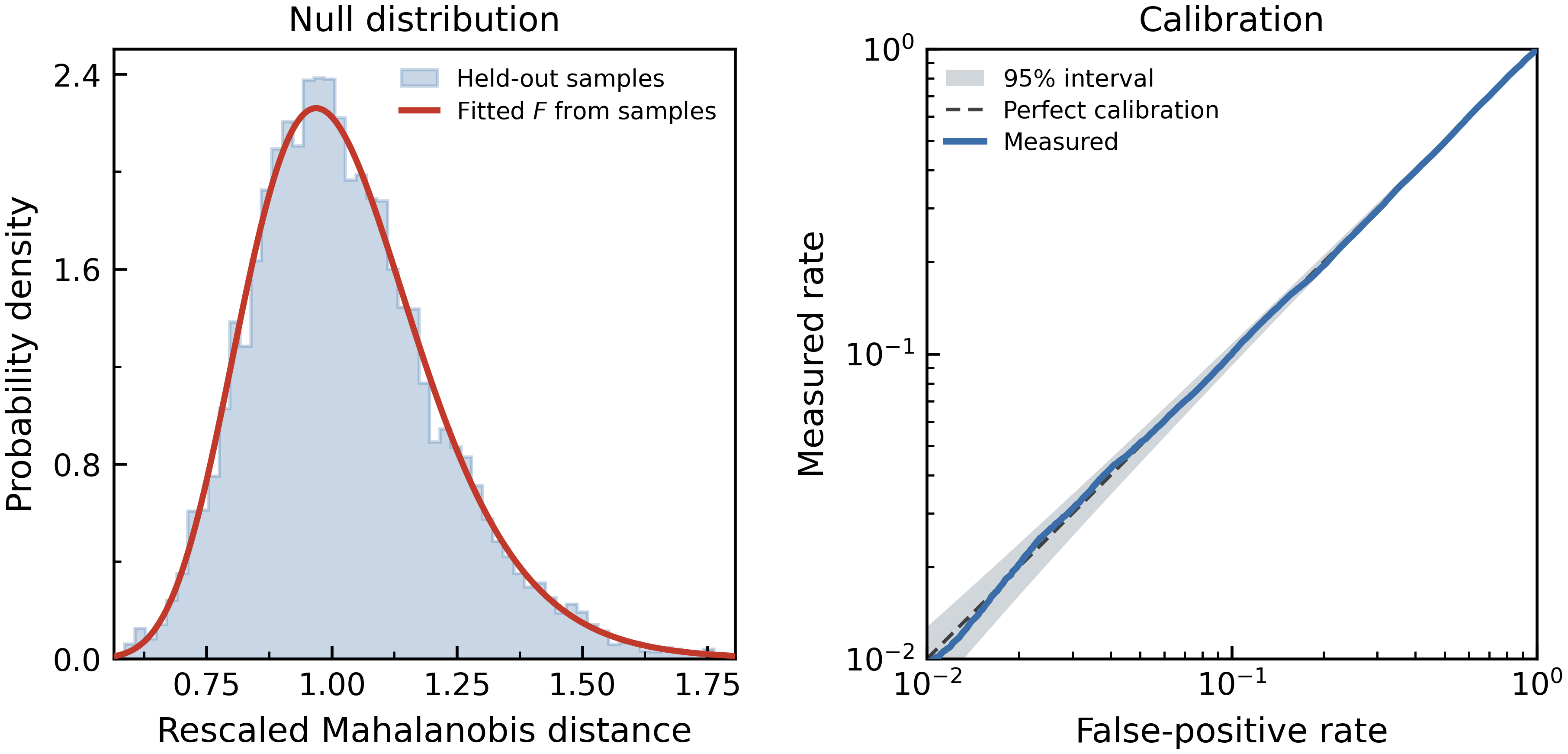}
    \caption{Check of the test's calibration on held-out reference simulations not used in the fit, for the \texttt{d0s0} case. \emph{Left:} rescaled Mahalanobis distances against the fitted $F$ distribution. \emph{Right:} measured rejection rate versus theoretical false-positive rate (blue), with the perfect-calibration diagonal (dashed) and its 95\% interval (grey).}
    \label{fig:calibration}
\end{figure}

\paragraph{Compatibility test} After the above reduction, each map is summarised by a feature vector $\mathbf{x}$, and we quantify the compatibility of \texttt{CMBsol} with a given Gaussian reference ensemble through the squared Mahalanobis distance
\begin{equation}
  T(\mathbf{x}) = (\mathbf{x} - \hat{\boldsymbol \mu})^{\top} \, \hat{\boldsymbol \Sigma}^{-1} \, (\mathbf{x} - \hat{\boldsymbol \mu}) \, ,
  \label{eq:mahalanobis}
\end{equation}
where $\hat{\boldsymbol \mu}$ and $\hat{\boldsymbol \Sigma}$ are the mean and covariance of the reference ST statistics; this measures the distance of \texttt{CMBsol} from the ensemble mean, weighted by the inverse covariance. Converting $T$ into a $p$-value requires knowing its distribution under the Gaussian null hypothesis, which would be a $\chi^2$ distribution for Gaussian features. However, the ST statistics of the reference maps are themselves non-Gaussian and heavier-tailed than a multivariate normal, even though the maps are Gaussian random fields. This effect thickens the tail of $T$, so a $\chi^2$ reference underestimates it and produces an overconfident, mis-calibrated test. We therefore calibrate the null distribution of $T$ empirically with an $F$ distribution fitted to the features of the reference Gaussian maps. We can then say whether a \texttt{CMBsol} is incompatible with the reference Gaussian maps when the $p$-value is lower than $0.05$. The calibration of the test is checked on held-out reference realisations that were not used to estimate $\hat{\boldsymbol \mu}$, $\hat{\boldsymbol \Sigma}$, or the fitted $F$ parameters. The left panel of figure \ref{fig:calibration} compares the distribution of their rescaled Mahalanobis distances with the fitted $F$ density, for the \texttt{d0s0} validation model: the two agree over the full range, including the upper tail that sets the $p$-values. The right panel shows the measured rejection rate as a function of the theoretical false-positive rate: it follows the diagonal within the 95\% interval expected from the finite number of held-out realisations, down to a false-positive rate of $10^{-2}$. The test is therefore correctly calibrated at the significance levels we use.

\section{Additional ST plots}
\label{sec:st_plots}
\begin{figure}
    \centering
    \includegraphics[width=1\textwidth]{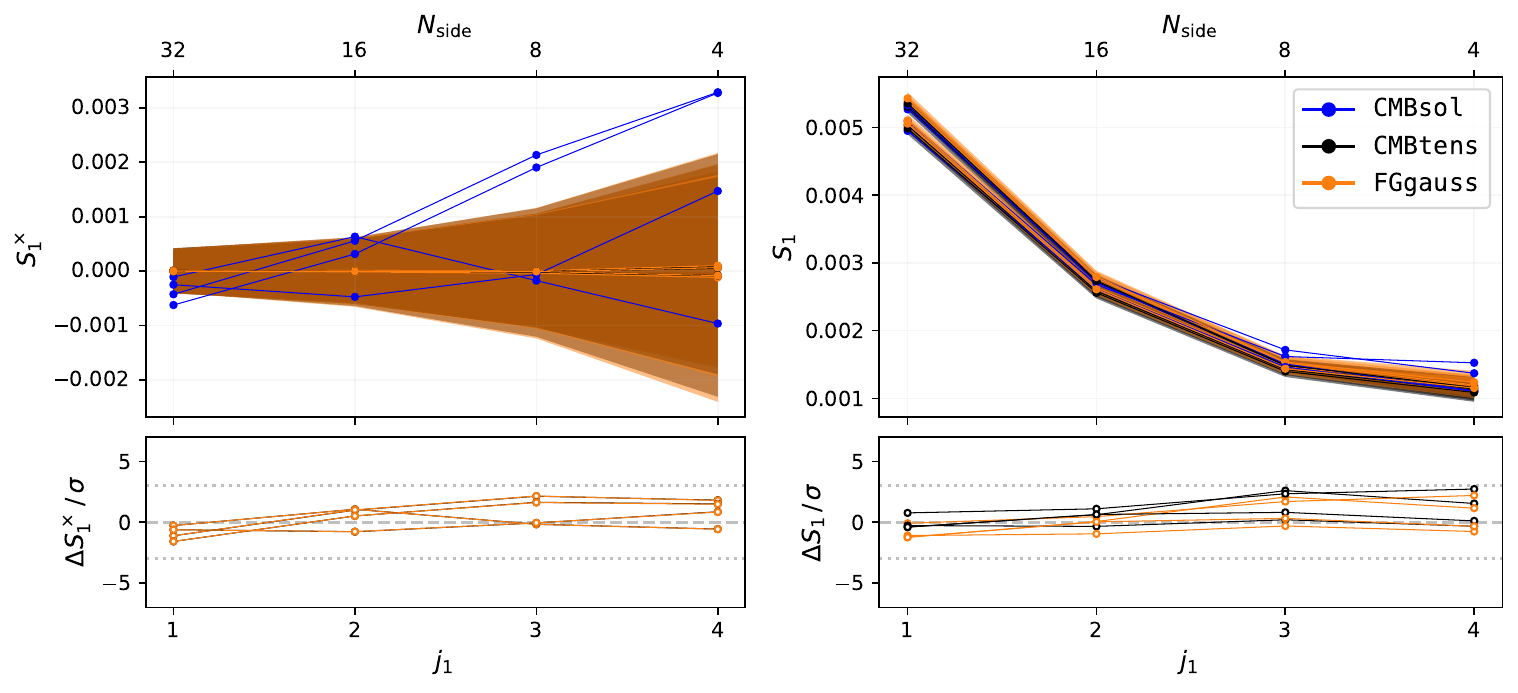}
    \caption{ST $S_1$ coefficient, same as figure \ref{fig:s1s2}, but for two validation cases. \emph{Left:} cross-coefficients correlating with synchrotron-dominated input data, for the MC-NILC - \texttt{d0s0} - $r_\text{in} = 0$ case. \emph{Right:} auto-coefficients, for the NILC - \texttt{d0s0} - $r_\text{in} = 5 \times 10^{-3}$ scenario. In both plots, the tensor-to-scalar ratio estimate is not biased, coefficients are compatible, and the robustness test raises no strong warnings.}
    \label{fig:s1s2_d0s0}
\end{figure}

In this section we report some additional plots of the scattering spectra for different scenarios. Figure \ref{fig:s1s2_d0s0} shows the $S_1$ and $S_2$ coefficients for two different validation cases discussed in section \ref{sec:results}. One plot is the same cross-synchrotron scenario of figure \ref{fig:s1s2}, but with the simpler foregrounds model \texttt{d0s0}. In the other one, the input $r$ is different from zero, simulating a true detection. In both cases, the input tensor-to-scalar ratio is recovered without bias, and the \texttt{CMBsol} coefficients are compatible with the Gaussian references ones. In these scenarios, the robustness test is passed and the detection is validated.

\section{Noise correlation}
\label{sec:noise_corr}
\begin{figure}
    \centering
    \includegraphics[width=0.5\textwidth]{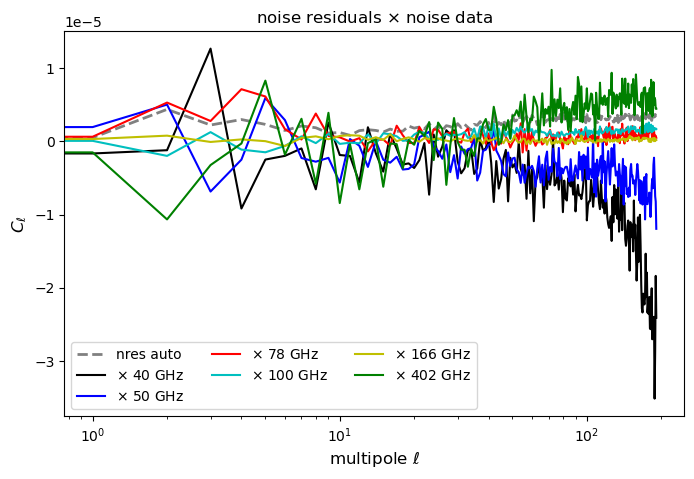}
    \includegraphics[width=0.45\textwidth]{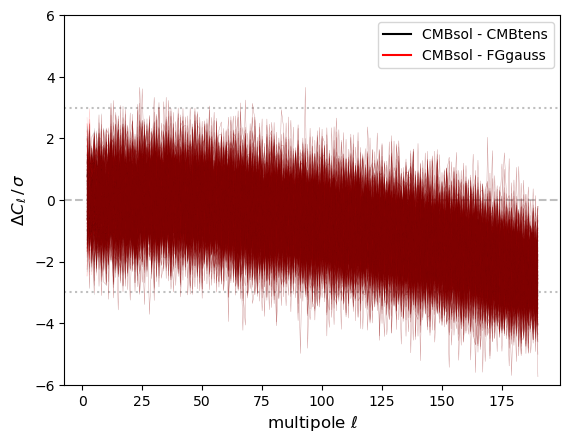}
    \caption{\emph{Left}: cross-power spectra between noise residuals and noise maps of different input frequency channels. \emph{Right}: difference between the cross-power spectrum of \texttt{CMBsol} maps and Gaussian simulations, in units of the standard deviation of the simulations $\sigma$, for the \texttt{d0s0} model. In this case, the measurement of $r$ is not biased by foreground residuals and compatibility is expected.}
    \label{fig:noise_corr}
\end{figure}

In section \ref{sec:test} we introduced the cross-ST coefficients, describing the correlation of \texttt{CMBsol} maps with foregrounds-dominated data. We highlighted one problem that may be encountered with this approach: since the component separation algorithm linearly combines the input multi-frequency data with some weights, the noise residuals in \texttt{CMBsol} are necessarily correlated with the noise of the input maps. The 40 and 402 GHz are the noisiest frequency channels, and the most correlated with the residuals, even if they are naturally down-weighted by the component separation method. This means that when we correlate with synchrotron- or dust-dominated maps, an extra contribution not due to foreground residuals is present both in the angular power spectrum and ST coefficients. This contribution is instead absent in the correlation with Gaussian simulations, as they are built with different noise realisations uncorrelated with the input data. The robustness test will thus register some false positives not induced by foregrounds residuals. This is illustrated in figure \ref{fig:noise_corr}: the left panel reports an example of the correlation between noise residuals and the noise in the input data for different frequencies, and for one simulation. In the right panel, we show the difference between the cross-power spectra of \texttt{CMBsol} and of the Gaussian simulations, correlated with 40 GHz data in the \texttt{d0s0} case, for all the 300 simulations. We notice that the 40 GHz channel is the one with the strongest noise correlation, and the same negative trend is present in both plots. The net impact of this contamination is a resulting efficiency of 100\% even for the \texttt{d0s0} model, where foreground residuals are completely mitigated.

In our case, this extra correlation is dominant at very small scales. To overcome this problem, we restrict the compatibility analysis to not include the smaller scales contaminated by this feature. For scattering spectra, we drop the $j = 0$ scale, while for angular power spectra the maximum multipole considered is $\ell_\text{max} = 95$\footnote{The scales considered for the two statistics are roughly the same, following the relation $N_\text{side} = 64 \cdot 2^{-j}$.}. We notice that in using this range of scales, we are still retaining some noise correlation in the summary statistics; we can not drop all the impacted scales as we would lose too much sensitivity for the robustness test. In any case, this just contributes in adding a few percent to the test efficiencies in the \texttt{d0s0} validation scenario (see tables \ref{tab:eta_j14} and \ref{tab:eta_cl_j14}), slightly raising the benchmark values above 10\%, but not changing the final conclusions of the study.

% Bibliography
\bibliographystyle{JHEP}
\bibliography{bsure2.bib}

\end{document}